\documentclass[a4paper,twocolumn,aps,prb,superscriptaddress,showpacs,floatfix]{revtex4}
\usepackage{amssymb,amsmath,stmaryrd}
\usepackage{times}

\usepackage{array}
\usepackage{graphicx}
\usepackage{ifpdf}
\usepackage{color}

\newcommand{\be}{\begin{equation}}
\newcommand{\ee}{\end{equation}}
\newcommand{\bea}{\begin{eqnarray}}
\newcommand{\eea}{\end{eqnarray}}

\def\a{\alpha}
\def\b{\beta}
\def\g{\gamma}
\def\G{\Gamma}
\def\d{\delta}
\def\D{\Delta}
\def\e{\epsilon}

\def\th{\theta}

\def\m{\mu}
\def\n{\nu}

\def\p{\pi}

\def\r{\rho}
\def\s{\sigma}
\def\S{\Sigma}
\def\t{\tau}

\def\vf{\varphi}

\def\w{\omega}

\def\q{\psi}
\def\Q{\Psi}

\def\bld{{\mathbf d}}
\def\ble{{\mathbf e}}

\def\bln{{\mathbf n}}

\def\blr{{\mathbf r}}

\def\blE{{\mathbf E}}

\def\blN{{\mathbf N}}

\def\callE{\mbox{$\mathcal{E}$}}

\def\callM{\mbox{$\mathcal{M}$}}

\def\callS{\mbox{$\mathcal{S}$}}

\def\callU{\mbox{$\mathcal{U}$}}

\def\bcallE{\mbox{\boldmath $\mathcal{E}$}}

\def\ua{\uparrow}
\def\da{\downarrow}
\def\ra{\rightarrow}

\def\iif{\infty}
\def\bra{\langle}
\def\ket{\rangle}

\def\Tr{{\rm Tr}}
\def\Re{{\rm Re}}
\def\Im{{\rm Im}}

\def\1op{\hat{\mathbbm{1}}}
\def\1{\mathbbm{1}}
\def\nn{\nonumber}

\begin{document}

\title{First-principles nonequilibrium Green's function approach to transient photoabsorption: 
Application to atoms}

\author{E. Perfetto}
\affiliation{Dipartimento di Fisica, Universit\`{a} di Roma Tor Vergata,
Via della Ricerca Scientifica 1, 00133 Rome, Italy}
\affiliation{INFN, Laboratori Nazionali di Frascati, Via E. Fermi 40, 00044 Frascati, 
Italy}

\author{A.-M. Uimonen}
\affiliation{Department of Physics, Nanoscience Center, FIN 40014, University of Jyv\"askyl\"a,
Jyv\"askyl\"a, Finland}

\author{R. van Leeuwen}
\affiliation{Department of Physics, Nanoscience Center, FIN 40014, University of Jyv\"askyl\"a,
Jyv\"askyl\"a, Finland; European Theoretical Spectroscopy Facility (ETSF)}

\author{G. Stefanucci}
\affiliation{Dipartimento di Fisica, Universit\`{a} di Roma Tor Vergata,
Via della Ricerca Scientifica 1, 00133 Rome, Italy; European Theoretical Spectroscopy Facility (ETSF)}
\affiliation{INFN, Laboratori Nazionali di Frascati, Via E. Fermi 40, 00044 Frascati, 
Italy}

\begin{abstract}
We put forward a first-principle NonEquilibrium Green's Function 
(NEGF) approach to calculate the
transient photoabsorption spectrum of optically thin systems. The method can deal with  
pump fields of arbitrary strength, frequency and duration as well as for 
overlapping and nonoverlapping pump and probe pulses. The 
electron-electron repulsion is accounted for by the correlation 
self-energy, and the resulting numerical scheme deals with 
matrices that scale quadratically with the 
system size. Two recent experiments, the first on helium and the 
second on 
krypton, are addressed. For the first experiment we explain
the bending of the Autler-Townes absorption peaks 
with increasing the pump-probe delay $\t$, 
and relate the bending to the thickness and density of the gas. For the second 
experiment we find that sizable spectral structures of the  
pump-generated admixture of Kr ions are fingerprints of 
{\em dynamical correlation} effects, and hence they cannot be 
reproduced by time-local 
self-energy approximations.
Remarkably, the NEGF approach also captures the retardation of the 
absorption onset of Kr$^{2+}$ with respect to Kr$^{1+}$ as a function 
of $\t$.
\end{abstract}


\maketitle

\section{Introduction}

Transient photoabsorption (TPA) spectroscopy has today become a popular technique 
to investigate the ultrafast dynamics of electrons and nuclei in 
atoms, molecules and 
nanostructures.\cite{ki.2009,bgk.2009,spsk.2012,ghlsllk.2013,kc.2014} A reliable physical 
interpretation of the 
TPA spectrum is inescapably linked to a reliable calculation of the 
probe-induced polarization in the pump-driven system. 
State-of-the-art calculations are based on the Configuration 
Interaction (CI) expansion of the time-evolved many-electron state.
The time-dependent CI coefficients are either varied with respect to the 
probe field\cite{gbts.2011,pghmetal.2011,cl.2012,tg.2012,psmwgs.2012,pbbmnl.2013} or  used to construct the 
dipole response function from the Lehmann 
representation,\cite{rs.2009,sypl.2011,blm.2012} the latter 
approach being applicable only provided that the dressed pump and 
probe fields do not overlap.\cite{ps.2015} However, the size 
of the arrays in CI calculations scale 
exponentially with the number of basis functions and the time-step to 
achieve convergence is typically much smaller than the 
time-step used in statistical approaches. Due to these numerical 
limitations the CI approach is confined to the study of rather small systems.

One possible statistical approach to TPA spectroscopy is Time Dependent Density 
Functional Theory (TDDFT).\cite{dbcwr.2013} In TDDFT  
are the occupied single-particle wavefunctions that are propagated in time 
and since they scale linearly 
with the number of basis functions TDDFT is suitable 
to study much larger systems than those accessible by CI.
In the framework of the Adiabatic Local Density Approximation (ALDA)  TDDFT
has been recently 
and successfully applied to the study of TPA in small- 
and medium-sized molecules\cite{nkyetal.2013} as well as to
monitor the vibronic-mediated charge transfer in 
donor-acceptor complexes.\cite{frbmetal.2014,rfsrmetal.2013} 
Still,  ALDA functionals have drawbacks that could compromise  
the description of photoabsorption spectra even in {\em equilibrium} 
systems. For example, 
ALDA misses correlation-induced spectral features
like double-excitations\cite{mzcb.2004,kk.2008} and 
long-range charge-transfer excitations;\cite{ngb.2006,m.2005,mt.2006} 
it also provides a poor description of the 
energy-level alignment in metal/molecule interfaces\cite{nhl.2006,srpss.2013} 
and of the Coulomb blockade phenomenon.\cite{sk.2011,ks.2013} In this 
work we discover yet another correlation effect  missed by ALDA.

An alternative statistical approach to TDDFT is the many-body 
diagrammatic theory. Here the building blocks of the formalism are 
the NonEquilibrium Green's Functions (NEGF),\cite{kb-book,hj-book,svl-book,bb-book}
and correlation effects 
are included by a proper selection of self-energy 
diagrams. Double-excitations and other properties missed by ALDA
are within reach of diagrammatic theory already with basic self-energies. 
Recently, it has been shown 
that the TPA spectrum follows from the solution of 
a nonequilibrium Bethe-Salpeter equation (BSE) provided that the 
time-scale of the pump-induced electron dynamics is much longer than 
the life-time of the dressed probe field.\cite{psms.2015} In general, 
however,
for pump fields of arbitrary strength, frequency and duration and/or 
for overlapping pump and probe pulses, the nonequilibrium BSE is inadequate and 
the full time-propagation of the NEGF is unavoidable. 
Nevertheless, 
as the size of the arrays in NEGF calculations  scales 
quadratically with the number of basis functions, this formalism too
allows for extending the range of CI accessible systems.
 
In this work we 
formulate a general first-principle NEGF scheme to TPA and apply it to 
reproduce the transient spectra of a thick helium gas\cite{pbbmnl.2013} and 
of a krypton gas.\cite{glwsetal.2010} For helium we address  
the exponential damping of the probe-induced dipole, and relate it to 
the thickness and density of the gas. We then
provide the explanation of the bending of the Autler-Townes absorption 
peaks with increasing the pump-probe delay. We also propose a useful formula 
for fitting the experimental TPA spectra. 
The krypton gas constitutes a  more severe test for NEGF due to
non-trivial correlation effects. We find that 
for a proper description of the  (pump-generated) evolving admixture of Kr 
ions the self-energy should have 
{\em memory}. Static (or adiabatic) approximations 
like the Hartree-Fock or the Markovian  
approximations perform rather poorly  as 
only the spectrum of Kr$^{1+}$ is visible.
Instead, the TPA spectrum calculated using the (memory-dependent) second-Born self-energy
contains absorption lines attributable to excitations in  
the Kr$^{1+}$ and Kr$^{2+}$ ions. Remarkably, we are also able to 
reproduce the femtosecond retardation of the absorption onset of Kr$^{2+}$ 
with respect to Kr$^{1+}$ as a function of the pump-probe delay.

The paper is organized as follows. In Section~\ref{tpasec} we relate 
the TPA spectrum to the microscopic quantum mechanical average of 
the transverse, probe-induced dipole-wave propagating toward the 
detector. In Section~\ref{negfsec} we put forward the NEGF approach 
to TPA and provide the explicit expression of the self-energy 
approximations implemented in this work. The helium gas is studied in 
Section~\ref{hesec}. In Section~\ref{opensec} we extend the NEGF approach 
to deal with pump-induced ionization processes
and then apply the extended approach to the study of the krypton 
gas in Section~\ref{Krsec}. Summary and conclusions are drawn in 
Section~\ref{concsec}.

\section{Transient photoabsorption spectrum}
\label{tpasec}

\begin{figure}[tbp]
\includegraphics[width=8.25cm]{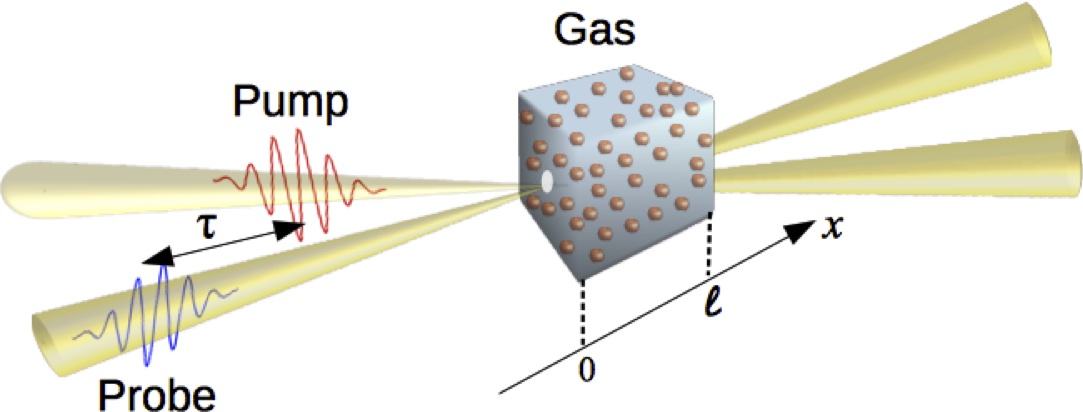}
\caption{(Color online) Schematic illustration of a pump-probe experiment.}
\label{ppill}
\end{figure}

We consider a gas of atoms or molecules perturbed by a strong time-dependent 
transverse electric field $\blE(\blr\,t)$ (pump) propagating along 
the unit vector $\blN$ and 
a feeble time-dependent transverse electric 
field $\ble(\blr\,t)$ (probe) propagating along 
the unit vector $\bln$, see Fig.~\ref{ppill}. 
The direction of propagation $\blN\neq \bln$ 
and in experiments the photodetector is positioned 
along the probe beam-line. 
Let $\bcallE_{p}(\blr\,t)$ be the component of the total electric field 
(external plus induced) propagating 
toward the detector and $\bcallE_{p}(t)$ be the value of 
$\bcallE_{p}(\blr\,t)$ at the detector surface. 
Then the transmitted energy measured by the detector is 
\be
E_{T}=\callS 
\frac{c}{2\p}\int\frac{d\w}{2\p}|\tilde{\bcallE}_{p}(\w)|^{2},
\label{ET}
\ee
with $\callS$ the surface of the sample (assumed to be smaller than 
the laser beam cross section).
Here and in 
the following we use the convention that quantities with the tilde symbol on top denote
the Fourier transform of the corresponding time-dependent
quantities. Replacing $\tilde{\bcallE}_{p}$ in 
Eq.~(\ref{ET}) with the external probe field $\tilde{\ble}$  we obtain the 
energy $E_{I}$ of the incident probe beam. The absorbed 
energy $E_{A}$ is defined according to
\bea
E_{A}&\equiv&E_{I}-E_{T}
\nn\\
&=&\callS 
\frac{c}{2\p}\int\frac{d\w}{2\p}\left(|\tilde{\ble}(\w)|^{2}-|\tilde{\bcallE}_{p}(\w)|^{2}\right).
\label{EA}
\eea
By means of a spectrometer it is possible to measure the absorbed 
energy per unit frequency, i.e.,
\be
\mathfrak{S}(\w)=\callS 
\frac{c}{2\p}\left(|\tilde{\ble}(\w)|^{2}-|\tilde{\bcallE}_{p}(\w)|^{2}\right).
\label{spectrum}
\ee
The quantity $\mathfrak{S}(\w)$ is the transient photoabsorption (TPA)
spectrum that we are interested in. 

As both $\blE$ and $\ble$ are transverse fields we have 
$\blE(\blr\,t)=\blE(X\,t)$  with 
$X=\blN\cdot\blr$ and $\ble(\blr\,t)=\ble(x\,t)$ with 
$x=\bln\cdot\blr$. By definition $\bcallE_{p}$  too
depends only on $x$, i.e., $\bcallE_{p}(\blr\,t)=\bcallE_{p}(x\,t)$, since it is a 
transverse field propagating along $\bln$.
Without loss of generality we choose the coordinates of the boundaries of the sample in 
$x=0$ and $x=\ell$, $\ell$ being the thickness  of the gas, see 
Fig.~\ref{ppill}. 
From Maxwell equations the relation between 
the total field and the incident probe field is
\be
\bcallE_{p}(x\,t)=\ble(x\,t)+\frac{2\p}{c}\frac{d}{dt}\int_{0}^{x}\!dx'\bld_{p}(x't)
\label{transe}
\ee
where $\bld_{p}(x\,t)$ is the component of the {\em probe-induced} 
dipole density
propagating toward the detector.\cite{ps.2015} Notice that 
$\bld_{p}(x\,t)$ vanishes for $x>\ell$; hence 
$\bcallE_{p}(\ell\,t)$ and $\bcallE_{p}(t)$ (the electric field at the 
detector surface) differ only by a time-shift. This time-shift is 
completely irrelevant for the calculation of the spectrum, so we can 
either use $\tilde{\bcallE}_{p}(\ell\,\w)$ or $\tilde{\bcallE}_{p}(\w)$ in 
Eq.~(\ref{spectrum}).

Equation~(\ref{transe}) connects the 
experimental outcome $\mathfrak{S}(\w)$ to a  
quantum-mechanical average. Let us define $\bld_{p}(x\,t)$ more 
rigorously. We denote by $|\Q(\blr\,t)\ket$ the many-body state of an atom 
located in a volume element around $\blr$ at time 
$t$ when both pump and probe fields are present; similarly  
$|\Q_{P}(\blr\,t)\ket$ is the many-body state of the same atom when only the 
pump is present (probe-free state). The average of 
the atomic dipole operator $\hat{\bld}(\blr)$ over these two states is 
$\bld^{(\rm at)}(\blr\,t)=\bra\Q(\blr\,t)|\hat{\bld}(\blr)|\Q(\blr\,t)\ket$ and 
$\bld^{(\rm at)}_{P}(\blr\,t)\equiv 
\bra\Q_{P}(\blr\,t)|\hat{\bld}(\blr)|\Q_{P}(\blr\,t)\ket$ respectively.  The 
atomic probe-induced dipole is therefore
\be
\bld_{p}^{(\rm at)}(\blr\,t)\equiv \bld^{(\rm 
at)}(\blr\,t)-\bld^{(\rm at)}_{P}(\blr\,t).
\label{probe-dip}
\ee
For isotropic systems the probe-free $\bld^{(\rm at)}_{P}(\blr\,t)=\bld^{(\rm at)}_{P}(X\,t)$ is a 
transverse field propagating along $\blN$. Instead the atomic dipole 
$\bld^{(\rm at)}(\blr\,t)$ is a transverse 
field propagating along all possible 
directions $k\bln+K\blN$ with $k$ and $K$ real numbers. To first 
order in $|\ble|$ the component propagating along $\blN$ ($k=0$) is exactly 
$\bld_{P}^{(\rm at)}(X\, t)$. It is reasonable to expect that for $k\neq 0$ and for small 
$|\ble|$ the dominant component of $\bld^{(\rm at)}(\blr\,t)$ is the one propagating 
along the same direction $\bln$ of the external probe. In this case 
the probe-induced atomic dipole $\bld^{(\rm at)}_{p}(\blr\,t)=\bld^{(\rm at)}_{p}(x\,t)$ is 
a transverse field propagating along $\bln$ and the function $\bld_{p}$ appearing in 
Eq.~(\ref{transe}) is simply $n^{(\rm at)}\bld^{(\rm at)}_{p}$, 
$n^{(\rm at)}$ being the density of the gas. We can therefore calculate $\bld_{p}$ by 
performing a time-propagation with pump and probe, another 
time-propagation with only the pump and then subtracting the resulting 
dipoles. Below we derive the basic equations to perform these calculations.

For an atom with unperturbed Hamiltonian $\hat{H}(\blr)$ the evolution of 
the
state $|\Q(\blr\,t)\ket$ is governed 
by the Schr\"odinger equation
\be
i\frac{d}{dt}|\Q(\blr\,t)\ket=\left[\hat{H}(\blr)+
\bcallE(\blr\,t)\cdot\hat{\bld}(\blr)\right]|\Q(\blr\,t)\ket
\label{seinr}
\ee
where $\bcallE$ is the total electric field. We choose 
the pump and probe propagation directions such that $\mathbf{N} \cdot \mathbf{n} \simeq 1$. 
Then $X=\blr\cdot\blN\simeq \blr\cdot\bln=x$ inside the sample and every $\blr$-dependent 
quantity can be approximated with an $x$-only dependent quantity. In 
particular the total field can be approximated as
\be
\bcallE(x\,t)=\blE(x\,t)+\ble(x\,t)+
\frac{2\p }{c}n^{(\rm 
at)}\frac{d}{dt}\int_{0}^{x}\!dx'\bld^{(\rm at)}(x'\,t).
\label{E+e+d}
\ee
For optically thin samples the dipole oscillations in two points $x_{1}$ and 
$x_{2}$ are time-shifted  by an amount $~|x_{1}-x_{2}|/c$ which is 
much smaller than the inverse of the typical dipole frequencies. 
Therefore $\bld^{(\rm at)}(x'\,t)\simeq \bld^{(\rm at)}(t)$ 
is weakly dependent on $x'$ inside 
the gas and we can approximate
the integral in Eq.~(\ref{E+e+d}) 
with  $x\bld^{(\rm at)}(t)$. 
The mathematical simplification brought about by this approximation
 is that every 
atom can be evolved separately. Consider the state $|\Q(t)\ket\equiv 
|\Q(\ell\,y\,z\,t)\ket$ of 
an atom at the interface in $x=\ell$ and let 
$\hat{H}\equiv\hat{H}(\ell\,y\,z)$  and 
$\hat{\bld}\equiv\hat{\bld}(\ell\,y\,z)$
be the atomic Hamiltonian and dipole operator. Then 
for $\blr=(\ell\,y\,z)$ Eq.~(\ref{seinr}) reads
\be
i\frac{d}{dt}|\Q(t)\ket=\left[\hat{H}+
\bcallE(t)\cdot\hat{\bld}\right]|\Q(t)\ket,
\label{seinl}
\ee
with
\be
\bcallE(t)=\blE(\ell\,t)+\ble(\ell\,t)+
\frac{2\p\ell}{c}n^{(\rm 
at)}\frac{d}{dt}\bld^{(\rm at)}(t)
\label{E+e+d-app}
\ee
and 
\be
\bld^{(\rm at)}(t)=\bra\Q(t)|\hat{\bld}|\Q(t)\ket.
\label{<d>}
\ee
Equations (\ref{seinl}-\ref{<d>}) form a close system of three 
coupled equations. The same equations could be solved for $\ble=0$ to 
obtain the probe-free atomic dipole 
$\bld^{(\rm at)}_{P}(t)$ and hence the probe-induced dipole 
$\bld^{(\rm at)}_{p}(t)$ in 
accordance with Eq.~(\ref{probe-dip}). Having  $\bld^{(\rm 
at)}_{p}(t)$ we can calculate the probe-induced dipole density 
$\bld_{p}(\ell\, t)=n^{(\rm at)}\bld^{(\rm at)}(t)$ and subsequently the component 
of the electric field propagating toward the detector from  
Eq.~(\ref{transe}), i.e.,
\be
\bcallE_{p}(\ell\,t)= 
\ble(\ell\,t)+\frac{2\p\ell}{c}\frac{d}{dt}\bld_{p}(\ell\, t).
\label{transe-simp}
\ee
Fourier transforming $\bcallE_{p}(\ell\,t)$ and $\ble(\ell\,t)$ 
the TPA spectrum $\mathfrak{S}(\w)$ in Eq.~(\ref{spectrum}) 
follows:
\be
\mathfrak{S}(\w)=-2\Im\left[\w\,\tilde{\ble}^{\ast}(\ell\,\w)\cdot\tilde{\bld}_{p}(\ell\,\w)\right]
-\frac{2\p\ell}{\callS c}\left|\w\,\tilde{\bld}_{p}(\ell\,\w)\right|^{2}.
\label{nedw}
\ee

The numerical solution of Eqs.~(\ref{seinl}-\ref{<d>}) is impractical
for heavy atoms or moderate size molecules. In the next section 
we describe a statistical approach which avoids solving the Schr\"odinger 
equation for the many-electron system. This approach is based on the Nonequilibrium 
Green's Functions\cite{kb-book,hj-book,svl-book,bb-book} (NEGF) 
combined with the Generalized Kadanoff-Baym Ansatz~\cite{lsv.1986} 
(GKBA), and it has been successfully applied to the
electron gas,~\cite{bksbkk.1996,kbbk.1998} two-band model 
semiconductors~\cite{hj-book,h.1992,bkbk.1997,bsh.1999,gbh.1999,vh.2000} 
and more recently bulk Silicon,~\cite{m.2013,sm.2015} 
Hubbard chains~\cite{bhb.2013,bhb2.2013,hsb.2014} and donor-acceptor 
junctions.~\cite{lpuls.2014} 
Here we extend it to perform first-principles 
simulations of laser-driven quantum systems.

\section{NEGF approach}
\label{negfsec}

We work in the formalism of second quantization and denote by 
$\hat{c}_{i\s}$ ($\hat{c}^{\dag}_{i\s}$) the annihilation 
(creation) operator for an electron in the orbital $\vf_{i}(\blr)$ 
with spin $\s=\ua,\da$. Without loss of generality the basis $\{\vf_{i}\}$ is 
taken orthonormal. The unperturbed Hamiltonian reads
\be
\hat{H}=\sum_{\substack{ij\\ \s}}h_{ij}\hat{c}_{i\s}^{\dag}\hat{c}_{j\s}+
\frac{1}{2}\sum_{\substack{ijmn\\ \s\s'}}
v_{ijmn}\hat{c}_{i\s}^{\dag}\hat{c}_{j\s'}^{\dag}\hat{c}_{m\s'}\hat{c}_{n\s}
\label{hamnano}
\ee
with $h_{ij}\equiv \int\! d\blr \,
\vf_{i}^{\ast}(\blr)[-\frac{1}{2}\nabla^{2}+V_{\rm 
n}(\blr)]\vf_{j}(\blr)$ the one-electron integrals with nuclear 
potential $V_{\rm n}$ and 
\be
v_{ijmn}\equiv \int\! d\blr\,d\blr'\,
\frac{\vf_{i}^{\ast}(\blr)\vf_{j}^{\ast}(\blr')\vf_{m}(\blr')\vf_{n}(\blr)}{|\blr-\blr'|}
\ee
the four-index Coulomb integrals. Similarly the dipole operator reads
\be
\hat{\bld}=\sum_{\substack{ij\\ \s}}\bld_{ij}\hat{c}_{i\s}^{\dag}\hat{c}_{j\s}
\ee
with $\bld_{ij}\equiv \int\! d\blr \,
\vf_{i}^{\ast}(\blr)\blr \vf_{j}(\blr)$ the matrix elements of the 
dipole vector. The time-dependent average of the atomic dipole, see 
Eq. (\ref{<d>}), is therefore 
\be
\bld^{(\rm at)}(t)=\sum_{\substack{ij\\ \s}}\bld_{ij}\bra\Q(t)|\hat{c}_{i\s}^{\dag}\hat{c}_{j\s}|\Q(t)\ket.
\label{<d>2}
\ee

Let us introduce the evolution operator $\callU(t,t_{0})$ from a time 
$t_{0}$ when the system is unperturbed to an arbitrary time $t$. Then 
$|\Q(t)\ket=\callU(t,t_{0})|\Q_{0}\ket$ with $|\Q_{0}\ket$ the state 
of the unperturbed system. To distinguish the 
Heisenberg from the Schr\"odinger picture we insert  a 
subscript ``$H$'' to an operator
$\hat{O}(t)$ in the 
Schr\"odinger picture,
$\hat{O}_{H}(t)\equiv\callU(t_{0},t)\hat{O}(t)\callU(t,t_{0})$. 
The lesser ($G^{<}$) and greater ($G^{>}$) Green's functions are 
defined according to 
\begin{subequations}
\bea
G^{<}_{ij}(t,t')=i\bra\Q_{0}|\hat{c}^{\dag}_{j\s,H}(t')\hat{c}_{i\s,H}(t)|\Q_{0}\ket,
\\
G^{>}_{ij}(t,t')=-i\bra\Q_{0}|\hat{c}_{i\s,H}(t)\hat{c}^{\dag}_{j\s,H}(t')|\Q_{0}\ket,
\eea
\label{g<g>}
\end{subequations}
and have the property 
\be
G^{\lessgtr}_{ij}(t,t')=-[G^{\lessgtr}_{ji}(t',t)]^{\ast},
\ee
implying that from $G^{\lessgtr}$ with times $t\geq t'$ or 
$t\leq t'$ we can reconstruct the entire $G^{\lessgtr}$. 
In this work we consider spin-compensated systems with no 
spin-orbit interaction; hence $G^{\lessgtr}$ takes the same value
for $\s=\ua,\da$. The NEGF formalism, however, is not 
limited to this case and it can easily be generalized to 
Hamiltonians that are not diagonal in spin space.
From $G^{<}$ at equal times we can calculate the 
time-dependent average of any one-body operator; in particular the 
atomic dipole in Eq.~(\ref{<d>2}) reads $\bld^{(\rm 
at)}(t)=-2i\sum_{ij}\bld_{ij}G^{<}_{ji}(t,t)$.

The lesser and greater Green's functions satisfy nonlinear 
integro-differential equations known as the Kadanoff-Baym 
equations (KBE).~\cite{kb-book,hj-book,svl-book,bb-book}
At present the numerical solution of 
the KBE for inhomogeneous systems is possible
only for moderate size basis 
sets,~\cite{dl.2007,mssvl.2008,mssvl.2009,bbvlsd.2009,fva.2009,bbb.2010,bbb2.2010} 
and still the CPU time is of the order of a few days for a 
propagation of $\sim 10^{3}$ time steps and $\sim 10^{1}$ basis functions. Considering that a TPA 
spectrum typically requires $10^{4}$ time steps for every delay 
between the pump and probe pulses, the  KBE 
are not feasible in this context.
A way to drastically reduce the computational effort 
consists in making the GKBA~\cite{lsv.1986} 
\begin{subequations}
\bea
G^{<}(t,t')&=&-G^{\rm R}(t,t')\r(t')+\r(t)G^{\rm A}(t,t'),
\label{gkba<}
\\
G^{>}(t,t')&=&G^{\rm R}(t,t')\bar{\r}(t')-\bar{\r}(t)G^{\rm A}(t,t').
\label{gkba>}
\eea
\label{gkba}
\end{subequations}
In these equations $G^{\rm R}(t,t')=[G^{\rm A}(t',t)]^{\dag}$ is the 
retarded Green's function and 
$\r(t)=-iG^{<}(t,t)=1-\bar{\r}(t)=1-iG^{>}(t,t)$
is the one-particle density matrix (the quantity of interest for the 
calculation of the atomic dipole). The GKBA is exact in the 
Hartree-Fock (HF) approximation and it is expected to be accurate 
in systems with well-defined quasiparticles, see 
Ref.~\onlinecite{lpuls.2014} for a recent discussion.
From the KBE we can easily derive, see Appendix~\ref{derivation}, the 
following equation of motion for $\r$ (in matrix form)
\be
-i\frac{d}{dt}\r(t)+\left[h_{\rm 
HF}(t),\r(t)\right]=iI(t)-{\rm H.c.},
\label{eomrho}
\ee
where the HF single-particle Hamiltonian reads
\be
h_{{\rm 
HF},ij}(t)=h_{ij}+\bcallE(t)\cdot\bld_{ij}+\sum_{mn}w_{imnj}\r_{nm}(t),
\label{HFham}
\ee
with
\be
w_{imnj}\equiv 2v_{imnj}-v_{imjn},
\ee
and the collision integral reads (in matrix form)
\be
I(t)=\!
\int_{-\iif}^{t}\! dt'\!\left[\S^{<}(t,t')G^{>}(t',t)+
\S^{>}(t,t')G^{<}(t',t)\right].
\label{collint}
\ee
The correlation self-energy $\S^{\lessgtr}$ appearing in Eq.~(\ref{collint}) is a functional of $G^{<}$ and 
$G^{>}$ which, in turn, are functionals of $\r$ through 
the GKBA. Thus, Eq.~(\ref{eomrho}) is a closed nonlinear 
integro-differential equation for $\r$ once 
we specify the functional form of the self-energy and the retarded 
Green's function. 

\begin{figure}[tbp]
\includegraphics[width=8.25cm]{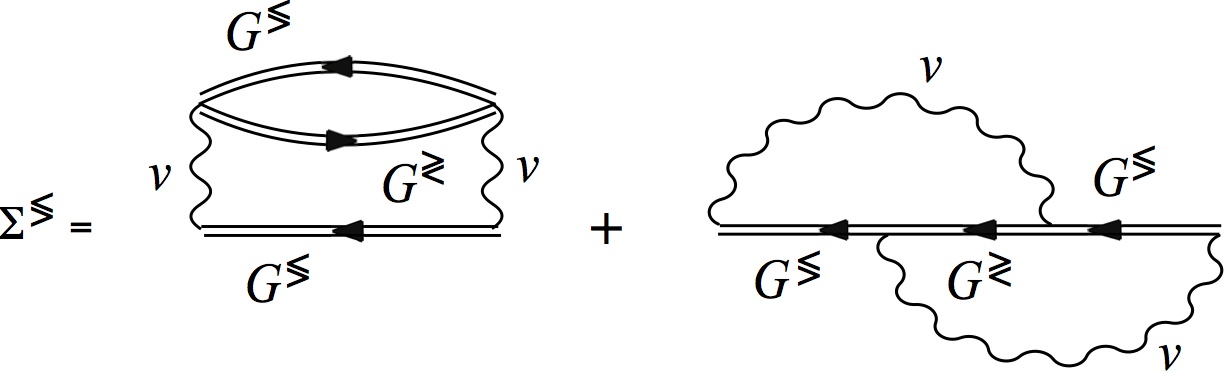}
\caption{Diagrammatic representation of the 2B self-energy. Wiggly 
lines denote the Coulomb interaction $v$.}
\label{2bsigma}
\end{figure}

In this work we present results obtained using the 
second-Born (2B) self-energy
\be
\S_{ij}^{\lessgtr}(t,t')=\!\!\!
\sum_{\substack{nm\\ pq\\sr}}v_{irpn}w_{mqsj}G_{nm}^{\lessgtr}(t,t')G_{pq}^{\lessgtr}(t,t')G^{\gtrless}_{sr}(t',t)
\label{2bse}
\ee
whose diagrammatic representation is shown in Fig.~\ref{2bsigma}.
For the retarded Green's function we consider 
\be
G^{\rm R}(t,t')=-i\th(t-t')\,T\left[e^{-i\int_{t'}^{t}d\bar{t}\,h_{\rm 
HF}(\bar{t})}\right]
\label{hfgr}
\ee
where $T$ is the time-ordering operator. In 
Ref.~\onlinecite{lpuls.2014} we discussed how to include correlation
effects (beyond HF) in $G^{\rm R}$ and showed their importance in quantum transport 
calculations. For the finite systems analyzed here, however, we found that 
the HF  $G^{\rm R}$ of Eq.~(\ref{hfgr}) is accurate enough. 
Similar findings were recently found  in strongly correlated 
systems as well.\cite{shbv.2015,lr.2015} In 
addition to the simplicity of Eq.~(\ref{hfgr}), the use of a HF  $G^{\rm R}$ within the 
NEGF+GKBA scheme guarantees the satisfaction of all basic 
conservation laws.~\cite{hsb.2014}

We emphasize that the equation of motion for $\r$ is an equation with 
memory since the evaluation of the collision integral at time $t$ 
involves the density matrix at all times $t'\leq t$. Using the GKBA 
expression for the lesser and greater Green's function the collision 
integral can be rewritten as
\begin{widetext}
\bea
I_{ik}(t)=\sum_{\substack{nm\\ pq\\sr}}\sum_{j}v_{irpn}w_{mqsj}\int_{-\iif}^{t} dt'
\left[
\left(G^{\rm R}(t,t')\r(t')\right)_{nm}\left(G^{\rm R}(t,t')\r(t')\right)_{pq}
\left(\bar{\r}(t')G^{\rm A}(t',t)\right)_{sr}
\left(\bar{\r}(t')G^{\rm A}(t',t)\right)_{jk}
\right.
\nn\\
+\left.
\left(G^{\rm R}(t,t')\bar{\r}(t')\right)_{nm}\left(G^{\rm R}(t,t')\bar{\r}(t')\right)_{pq}
\left(\r(t')G^{\rm A}(t',t)\right)_{sr}
\left(\r(t')G^{\rm A}(t',t)\right)_{jk}\right].
\label{collgkba}
\eea
\end{widetext}
From Eq.~(\ref{collgkba}) it is evident that the computational cost scales 
quadratically with the maximum propagation time. 
The quadratic scaling can be reduced to a linear scaling if 
the collision integral is evaluated in the 
Markov approximation. The Markov approximation 
consists in neglecting memory effects by replacing the pair $\r(t')$ and 
$\bar{\r}(t')$ with the pair $\r(t)$ and 
$\bar{\r}(t)$, and in using the  {\em equilibrium} HF retarded Green's 
function $G^{\rm 
R}(t,t')=-i\th(t-t')\exp[-ih_{\rm HF}^{\rm eq}(t-t')]$, with 
$h_{{\rm HF}}^{\rm eq}$ 
the HF single-particle Hamiltonian of the 
equilibrium system. In this case the integral over $t'$ can be done 
analytically and the collision integral becomes a quartic polynomial 
in $\r(t)$. Thus the equation of motion for $\r$ reduces to a  
nonlinear differential equation. In the 
next sections we benchmark the Markov approximation against 
Configuration Interaction (CI) and full NEGF+GKBA simulations.

\section{Helium}
\label{hesec}

In this section we simulate the TPA spectrum of helium atoms recently measured in 
Ref.~\onlinecite{pbbmnl.2013}. Helium is among one of the most 
studied systems in 
the context of TPA~\cite{gbts.2011,rthzetal.2011,cl.2012,pl.2012,cbbmetal.2012,pbbmnl.2013,cwgs.2013,wcgs.2013,cwcc.2014} 
and, due to the limited number of electrons, the CI simulations
are very accurate. From our point of view helium provides
an extremely useful platform to benchmark  the NEGF+GKBA methodology.  
The gas of He atoms is perturbed by a 
near infrared (NIR) transverse pump pulse $\blE(\ell \,t)=(E(t),0,0)$ with 
$E(t)=E_{0}\sin^{2}(\p t/\D_{P})\sin(\w_{P}t)$ for $0<t<\D_{P}$. The 
experimental pump intensity is $\mathfrak{I}_{0}=6\times 10^{12}$ 
W/cm$^{2}$, which 
corresponds to an electric field 
$E_{0}=\sqrt{2\mathfrak{I}_{0}/c\e_{0}}=6.6\times 10^{9}$ V/m, the 
duration of the pump pulse is $\D_{P}\sim 15$  fs and the NIR 
frequency, $\w_{P}\sim 0.57$ eV, is slightly detuned from the $2s-2p$ 
resonance. 
Thus, the pump alone does not perturb the equilibrium state of the He 
atoms as both the $2s$ and $2p$ levels are empty. The situation changes 
if the probe pulse arrives first. In the experiment the probe field 
is an ultrashort pulse $\ble(\ell \,t)=(e(t),0,0)$, 
with $e(t)=e_{0}\sin^{2}(\p (t-\tau)/\D_{p})\sin(\w_{p}(t-\t))$ for 
$\t<t<\t+\D_{p}$. The probe pulse has duration
$\D_{p}\sim 0.5$ fs, it is centered at frequency $\w_{p}=22$ eV and it 
has an intensity $\mathfrak{i}_{0}\sim  10^{9}$ 
W/cm$^{2}$, which corresponds to an electric field $e_{0}=8.6\times 
10^{7}$ V/m. The density $n^{(\rm at)}$ can been deduced from the 
pressure $P$ of the gas: $n^{(\rm at)}=P/(K_{B}T)$ with $K_{B}$ the Boltzmann 
constant. The experimental estimate of $P$ varies in the range 
$10-240$ mbar, implying that  at room temperature $n^{(\rm at)}$ varies in the range 
$5.8\times 10^{16} - 2.4\times 10^{17}$ cm$^{-3}$. 
Finally we observe that the experimental thickness (1 $\m$m $\div$ 1 mm) is much larger 
than the wavelength of the laser pulses (optically thick samples). 
To deal with these thicknesses, we propose the following approximation
$\int_{0}^{\ell}dx' \bld^{\rm (at)}(x'\,t)\simeq 
\ell_{\rm eff}\bld^{\rm (at)}(\ell\,t)$ where the effective thickness $\ell_{\rm eff}<\ell$ 
can be used as a fitting parameter. 
As we shall see this approximation well captures the  
effects of screening in the TPA spectrum of a thick helium gas.

We obtained the one- and two-electron 
integrals as well as  the dipole matrix elements from the SMILES 
package~\cite{smiles1,smiles2} using the VB2 Slater-type orbital (STO)
basis, consisting of 15 basis functions. We performed CI 
time-propagations as well as NEGF+GKBA propagations 
in the HF ($\S=0$), 2B and Markovian 2B approximation. As a general comment we 
observe that the time-step to achieve convergence in CI is about an 
order of magnitude smaller than in NEGF+GKBA, thereby CI  
requires about ten-times more time steps than NEGF+GKBA to have the same 
frequency resolution.

\begin{figure}[tbp]
\includegraphics[width=4.25cm]{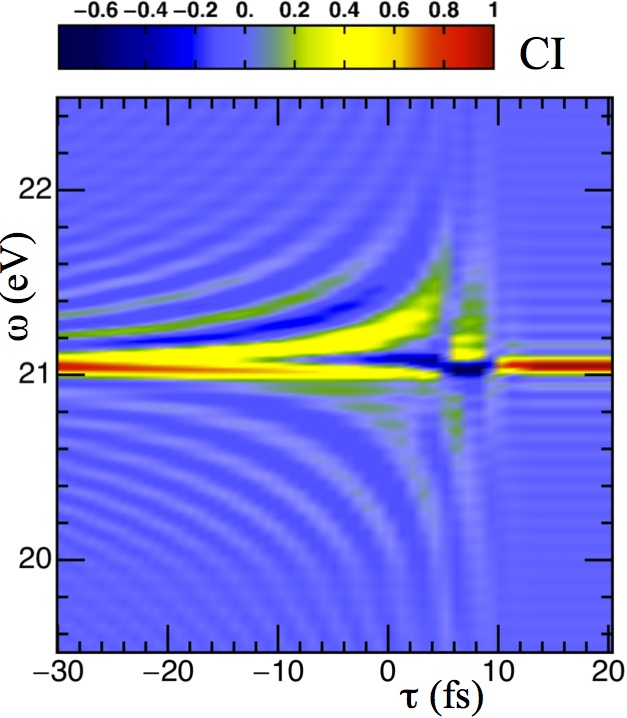}
\includegraphics[width=4.29cm]{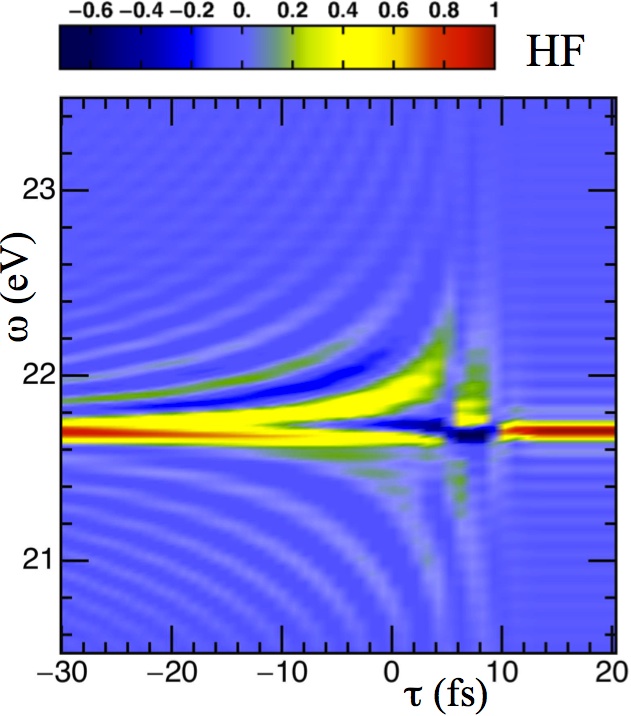}
\includegraphics[width=4.25cm]{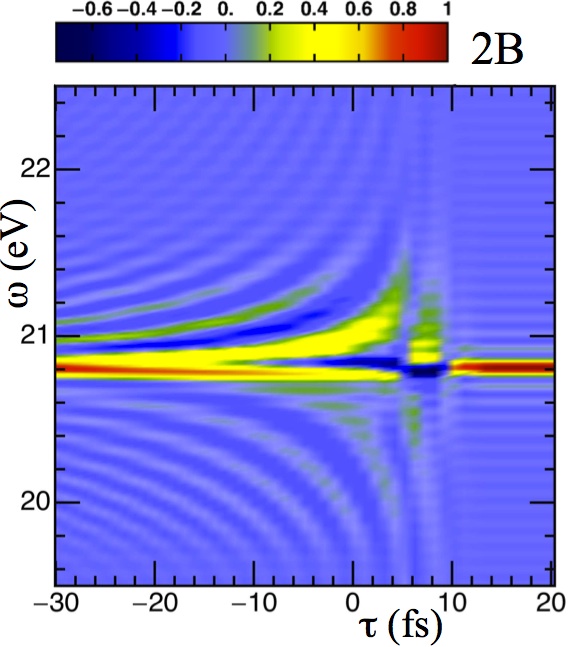}
\includegraphics[width=4.25cm]{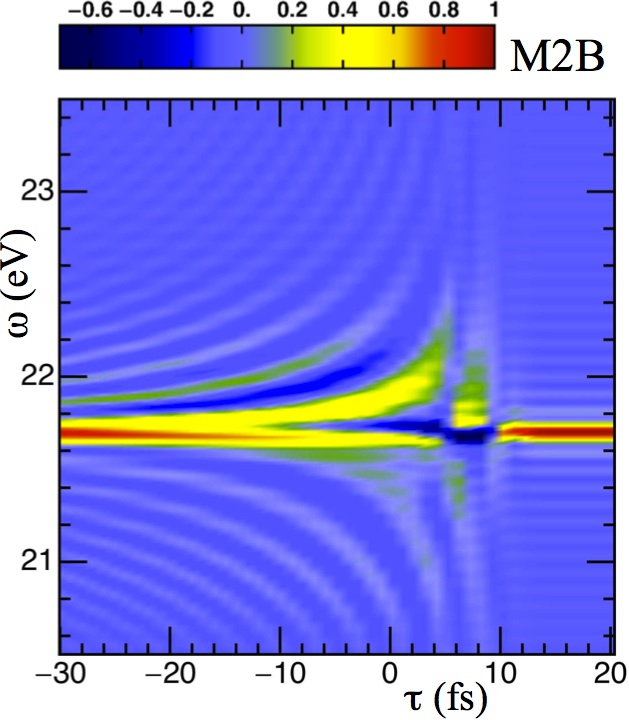}
\caption{(Color online) Density plot of the TPA spectrum (normalized to the maximum height) 
of a helium gas with $\ell_{\rm eff}=0.16$~mm and density $n^{(\rm at)}=2.4\times 
10^{17}$~cm$^{-3}$. The pump and probe pulses are given in the main 
text. Results obtained with CI (top-left), HF (top-right), 2B 
(bottom-left) and the Markovian 2B (bottom-right).}
\label{He-thk0005}
\end{figure}

In Fig.~\ref{He-thk0005} we compare the TPA spectrum in the four 
different schemes for a 
gas with $\ell_{\rm eff}=0.16$~mm and density $n^{(\rm at)}=2.4\times 
10^{17}$~cm$^{-3}$. The spectra are obtained by a fast Fourier 
transform of $\bld_{p}^{\rm (at)}(t)$ calculated using a time-step $\D t=0.0012$ fs and 
40.000 time steps in CI, and $\D t=0.006$ fs and 7.500 time steps
in NEGF+GKBA. In CI, see top-left panel, the equilibrium peak of the 
$1s-2p$ transition occurs at frequency $21.1$~eV, in agreement 
with the experiment, whereas the $1s-3p$ transition is not accurate.
Nevertheless, as our interest is in the evolution of the $1s-2p$ 
transition as the pump-probe delay is varied,  
the VB2 basis is suitable for our purposes. The overall shape of the 
CI TPA spectrum is well reproduced by 
all NEGF approximations, see, e.g., the size of the splitting and
the relaxation toward the equilibrium spectrum. The only relevant 
difference is a uniform frequency 
shift. 
We also observe that the intensity of the 
equilibrium peak grows monotonically with decreasing $\t$ (no coherent 
oscillations), a feature in common with the experiment. 
In HF, see top-right panel, the energy of the $1s-2p$ transition is overestimated. 
The inclusion of correlation effects at the 2B level, see bottom-left panel, 
counteracts this overestimation, although the correction is too large. 
Interestingly the Markovian 2B approximation, see  bottom-right panel, does not change the 
position of the HF peak,  a result which points to the importance of memory 
effects, see also Section~\ref{Krsec}.

\begin{figure}[tbp]
\includegraphics[width=4.25cm]{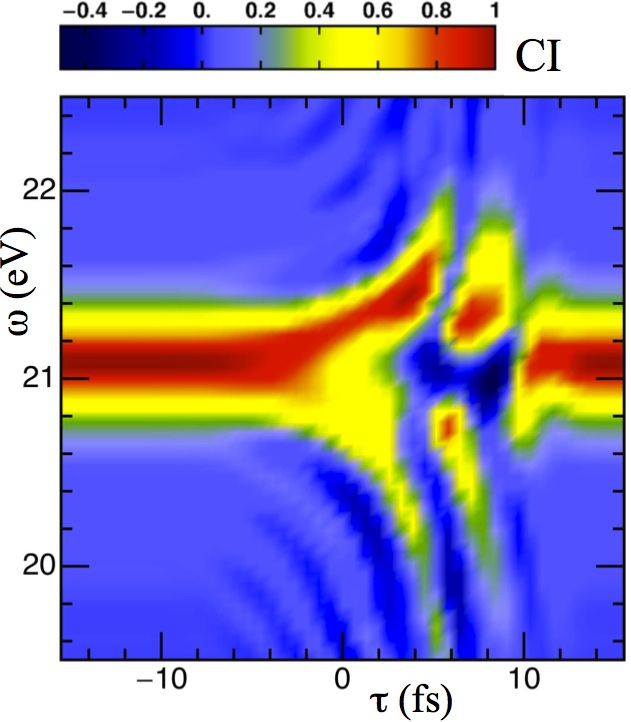}
\includegraphics[width=4.25cm]{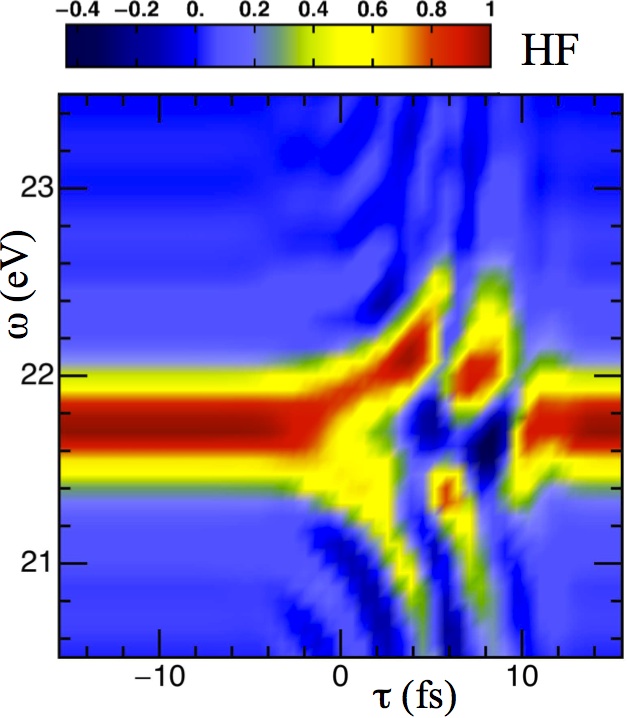}
\includegraphics[width=4.25cm]{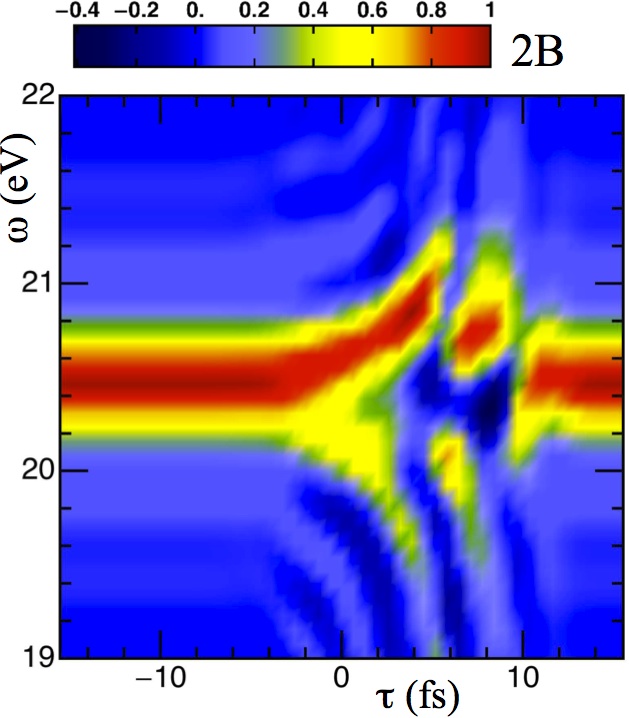}
\includegraphics[width=4.25cm]{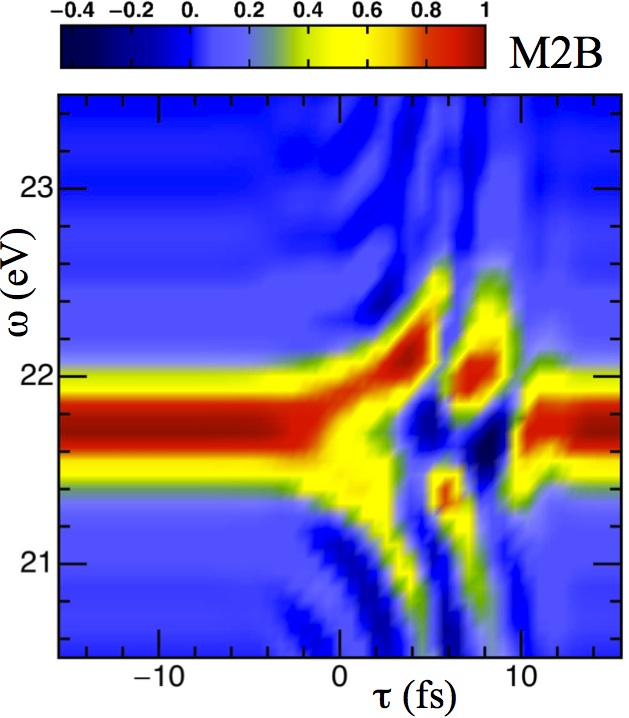}
\caption{(Color online) Density plot of the TPA spectrum (normalized to the maximum height) 
of a helium gas with $\ell_{\rm eff}=1.6$~mm and density $n^{(\rm at)}=2.4\times 
10^{17}$~cm$^{-3}$. The pump and probe pulses are given in the main 
text. Results obtained with CI (top-left), HF (top-right), 2B 
(bottom-left) and the Markovian 2B (bottom-right).}
\label{He-thk005}
\end{figure}

The performance of the NEGF+GKBA approach is satisfactory for larger 
thicknesses too. In Fig.~\ref{He-thk005} we compare the TPA spectrum for  
$\ell_{\rm eff}=1.6$~mm and density $n^{(\rm at)}=2.4\times 
10^{17}$~cm$^{-3}$ in 
the four schemes. The fast Fourier 
transform of $\bld_{p}^{\rm (at)}(t)$ has been performed with $\D t=0.0012$ fs and 
10.000 time steps in CI, and $\D t=0.006$ fs and 1.500 time steps
in NEGF+GKBA. Again all main features of the
CI spectrum are well reproduced.
The width of the absorption peaks is larger 
as compared to Fig.~\ref{He-thk0005} since the life-time of the 
probe-induced dipole is about an order of magnitude shorter.

Let us now come to the physical interpretation of the TPA spectrum. 
For small $\t$ the equilibrium peak 
of the $1s-2p$ transition undergoes an Autler-Townes (AT) splitting 
since the $2p$ and $2s$ levels are mixed by the pump field. 
The asymmetry of the AT intensities is due to the fact that
$\w_{P}$ is not exactly tuned at the $2s-2p$ resonance.  
For a pump of finite duration to generate 
an AT splitting the probe-induced 
dipole {\em must} decay in a time-window 
$W\lesssim \D_{P}$. In fact, the pump affects the  oscillations 
of $\bld_{p}(t)$ only in a 
time-window $\D_{P}$, hence it cannot change the position of the 
peaks of $\tilde{\bld}_{p}(\w)$ if $W\gg \D_{P}$.\cite{ps.2015} 
As the time evolution is unitary and the system is 
finite the damping mechanism of the dipole moment is not driven by 
electron-electron scattering. The damping mechanism does not have its 
origin in the radiative decay either (in He relaxation through 
radiative decay is no shorter than hundreds of fs). We will explain 
the origin of the damping mechanism in Section~\ref{dampingsec}. 
Here we address a different issue, i.e., the bending of the AT 
absorption peaks as $\t$ decreases and the possible formation  of a subsplitting structure.

Consider a three-level He model with basis functions $1s$, $2s$, $2p$ in the
oscillating state induced by the probe. If the probe is switched off 
at $t=0$ then for $t>0$ the probe-induced dipole (along the $x$ component) is 
$d_{p}(t)=d_{0}\sin(\w_{0}t)$, where $\w_{0}$ is the energy 
of the $1s-2p$ transition and $d_{0}=d_{x,1s2p}$ is the dipole matrix 
element. At time $-\t>0$ we switch on a pump field 
$E(t)=E_{0}e^{-\g_{P}(t+\t)}\sin(\w_{P}(t+\t))$,
of duration $\D_{P}\sim 1/\g_{P}$, which couples the $2s$ and $2p$ levels. 
We choose $\w_{P}$ in resonance with the energy of the $2s-2p$ 
transition and work in the rotating wave approximation. Then for times 
$t>-\t$ we find~\cite{ps.2015} 
$d_{p}(t)=d_{0}\sin(\w_{0}t)\cos\!\left(\!E_{0}d_{0}\frac{1-e^{-\g_{P}(t+\t)}}{\g_{P}}\right)$ (we 
assumed for simplicity that the matrix element $d_{x,2s2p}= d_{0}$). Collecting 
these results and introducing an exponential damping $\g\sim 
1/W$ (the 
origin of which is explained in Section~\ref{dampingsec}) we can 
write 
\be
\frac{d_{p}(t)}{d_{0}}=e^{-\g t}\sin(\w_{0}t)
\times\left\{
\begin{array}{lr}
    \!\!1 & t<-\t \\
\!\!\cos\!\left(\!E_{0}d_{0}\frac{1-e^{-\g_{P}(t+\t)}}{\g_{P}}\right)
 & t>-\t
 \end{array}
\right.
\label{dipmodel}
\ee
and $d_{p}(t)=0$ for $t<0$ (before the probe). 
This equation clearly illustrates the behavior 
previously discussed. For 
$t>-\t+1/\g_{P}$ the cosine is essentially constant. Thus, if 
$\g\ll\g_{P}$ (hence $W\gg\D_{P}$) then the pump modifies
the $\sin(\w_{0}t)$ profile only in the time window 
$(-\t,-\t+1/\g_{P})$, too short to change the position of the peaks 
in $\pm\w_{0}$ of the Fourier transform $\tilde{d}_{p}(\w)$. Let us now consider the 
opposite limit $\g\gg\g_{P}$ (hence  $W\ll\D_{P}$). For all times $t$
smaller than $1/\g$ (after this time the dipole is exponentially 
small) we can approximate the cosine with $\cos(E_{0}d_{0}(t+\t))$. 
In this approximation the Fourier transform $\tilde{d}_{p}(\w)$ has a simple 
analytic form and for, e.g., $\w\simeq \w_{0}+E_{0}d_{0}$, we find
\be
-\Im[\tilde{d}_{p}(\w)]\simeq\frac{e^{-\g\t}}{4}
\frac{\g\cos(\bar{\w}\t)+(\bar{\w}-E_{0}d_{0})\sin(\bar{\w}\t)}{(\bar{\w}-E_{0}d_{0})^{2}+\g^{2}}
\label{ATtau}
\ee
with $\bar{\w}\equiv \w-\w_{0}$. The denominator of Eq.~(\ref{ATtau}) 
has a minimum in $\bar{\w}=E_{0}d_{0}$ which is {\em independent} of $\t$. 
However, the maximum of $-\Im[\tilde{d}_{p}(\w)]$ does not coincide with 
the minimum of the denominator as $\t$ decreases from zero. For small 
negative $\t$ the maximum occurs at frequencies $\bar{\w}\simeq 
E_{0}d_{0}(1+\t\g/2)<E_{0}d_{0}$. This explains the bending of the AT 
absorption peaks 
(a similar analysis can be done for frequencies $\w\simeq 
\w_{0}-E_{0}d_{0}$). 

\begin{figure}[tbp]
\includegraphics[width=4.29cm]{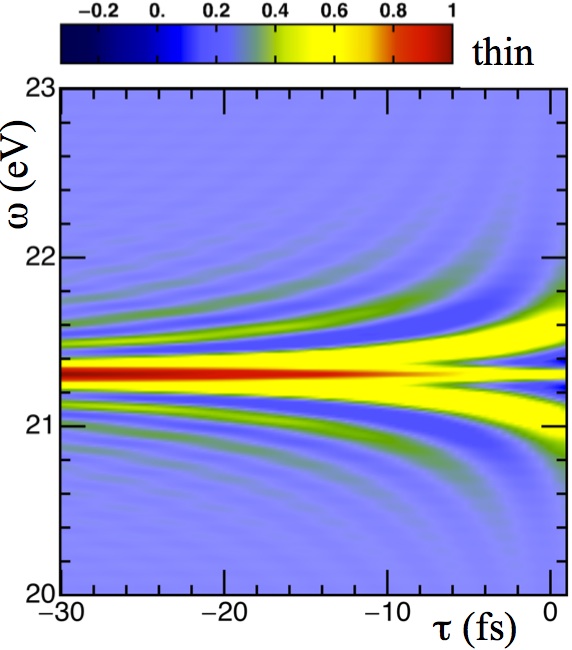}
\includegraphics[width=4.25cm]{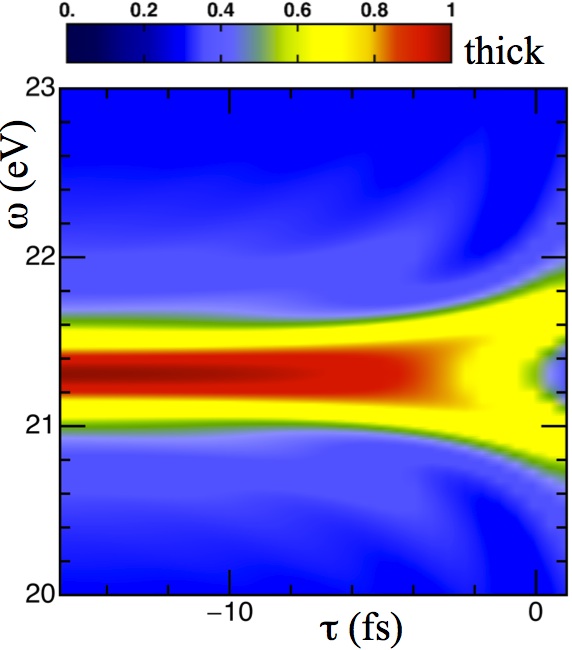}
\caption{(Color online) Density plot of $-\w\Im[\tilde{d}_{p}(\w)]/d_{0}$, 
with $\tilde{d}_{p}$ the Fourier transform of the function $d_{p}$ in 
Eq.~(\ref{dipmodel}), for $\w_{0}=21.3$~eV, $E_{0}d_{0}=0.5$~eV,
$\g_{P}=0.41$ eV and for $\g=0.21$ eV (left panel) and  $\g=1.37$ eV 
(right panel).}
\label{toyfig}
\end{figure}

\begin{figure}[tbp]
\includegraphics[width=0.5\textwidth]{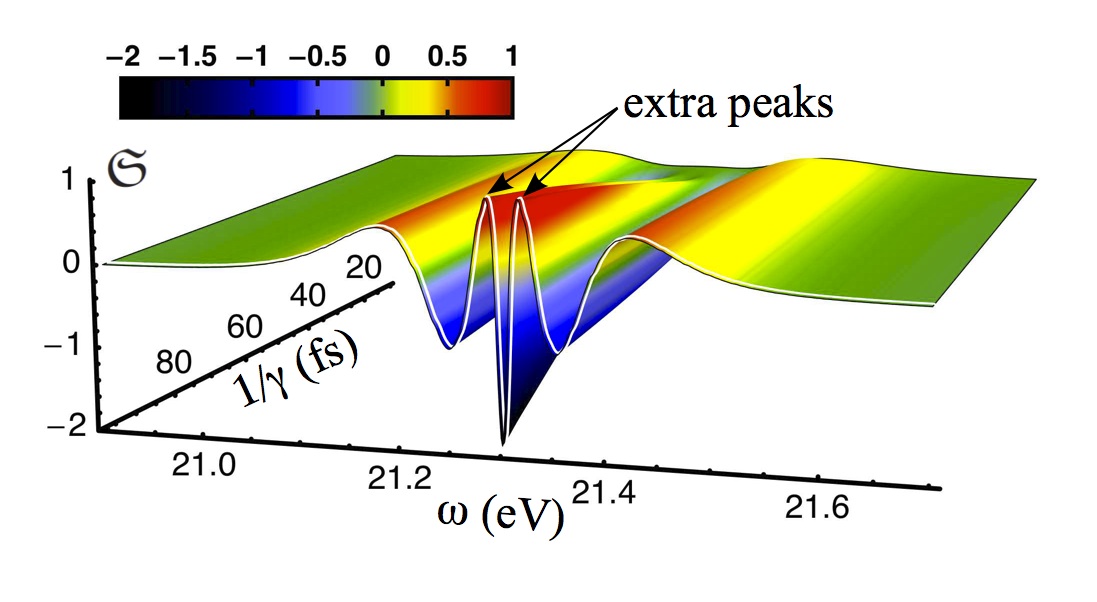}
\caption{(Color online) 3D plot of $-\w\Im[\tilde{d}_{p}(\w)]/d_{0}$ (normalized to 
the maximum height), 
with $\tilde{d}_{p}$ the Fourier transform of the function $d_{p}$ in 
Eq.~(\ref{dipmodel}),  as a function of the dipole life time $1/\g$ at delay $\t=0$ for $\w_{0}=21.3$~eV, 
$E_{0}d_{0}=0.2$~eV, $\g_{P}=0.125$ eV. Arrows indicate the extra peaks discussed in the main text.}
\label{subsplit}
\end{figure}

Noteworthy, the full Fourier transform of the 
dipole in Eq.~(\ref{dipmodel}) yields spectra that resemble very 
closely those in Figs.~\ref{He-thk0005} and~\ref{He-thk005}. 
In Fig.~\ref{toyfig} we 
display the density plot of $-\w\Im[\tilde{d}_{p}(\w)]\propto 
\mathfrak{S}(\w)$ for $\w_{0}=21.3$~eV, $E_{0}d_{0}=0.5$~eV,
$\g_{P}=0.41$ eV (corresponding to a duration $\D_{P}\sim 10$ fs) 
and for $\g=0.21$ eV (left panel, compare with 
Fig.~\ref{He-thk0005}) and
$\g=1.37$ eV (right panel, compare with Fig.~\ref{He-thk005}).
Equation~(\ref{dipmodel}) does therefore provide a convenient 
analytic parametrization of experimental TPA spectra. For longer 
pumps Eq.~(\ref{dipmodel}) predicts the formation of a subsplitting 
structure as well. In Fig.~\ref{subsplit} we show the TPA spectrum as 
a function of the dipole life-time $1/\g$ at 
delay $\t=0$ for $\w_{0}=21.3$~eV, $E_{0}d_{0}=0.2$ eV, 
$\g_{P}=0.125$ eV (corresponding to a duration $\D_{P}\sim 33$ fs). 
In addition to the AT peaks  at $\w_{0}\pm 
E_{0}d_{0}$ two extra peaks emerge, in agreement with recent 
experiments on a thick helium gas perturbed by a NIR pump 
of duration 33 fs.\cite{lscsg.2015} We point out that  
the number of extra peaks increases with increasing the AT splitting 
(hence with increasing the intensity of the NIR pulse), a prediction which could 
be easily checked experimentally.

\subsection{Damping in closed systems with unitary evolution}
\label{dampingsec}

The damping of the probe-induced dipole in the helium gas 
has been systematically investigated both numerically and 
experimentally in Ref.~\onlinecite{lscsg.2015}. Here we 
provide a transparent explanation based on the analytic solution of  
the Schr\"odinger equation~(\ref{seinl}) for a 
simple two-level model with two-particle states $|1s^{2}\ket$ and 
$|1s2p_{x}\ket$. We take the equilibrium Hamiltonian $\hat{H}$ diagonal 
on this basis and let $\e_{1}$ and $\e_{2}$ be the corresponding 
eigenenergies. We denote by $d_{0}=\bra 
1s^{2}|\hat{d}_{x}|1s2p_{x}\ket$ the $x$-component of dipole matrix 
element and write the dressed probe field along $x$ in accordance 
with Eq.~(\ref{transe-simp}), i.e., 
$\callE_{p}(t)=e(t)+\a\dot{d}(t)$, where $\a=2\p\ell_{\rm eff} n^{(\rm at)}/c>0$ 
and $d(t)=\bra\Q(t)|\hat{d}_{x}|\Q(t)\ket$. Expanding the 
time-dependent two-particle state as $|\Q(t)\ket = 
c_{1}(t)|1s^{2}\ket+c_{2}(t)|1s2p_{x}\ket$ we find
\begin{subequations}
\bea
i\dot{c}_{1}(t)&=&\e_{1}c_{1}(t)+d_{0}\callE_{p}(t)c_{2}(t),
\\
i\dot{c}_{2}(t)&=&\e_{2}c_{2}(t)+d_{0}\callE_{p}(t)c_{1}(t).
\eea
\end{subequations}
For any real $\callE_{p}$ the time evolution of the coefficients 
$c_{1}$ and $c_{2}$ is unitary. 
The expression of the time-dependent dipole in terms of $c_{1}$ and $c_{2}$ reads 
$d(t)=2d_{0}\Re[c_{1}^{\ast}(t)c_{2}(t)]$. We can get a differential 
equation for $d$ if we introduce two more real functions 
$g(t)=\Im[c_{1}^{\ast}(t)c_{2}(t)]$ and 
$f(t)=|c_{1}(t)|^{2}-|c_{2}(t)|^{2}$. It is a matter of simple 
algebra to show that after the probe (hence $e(t)=0$) these three functions 
satisfy the system
\begin{subequations}
\bea
\dot{f}(t)&=&-2d_{0}\a g(t)\dot{d}(t),
\label{hdot}
\\
\dot{g}(t)&=&-\frac{\w_{0}}{2d_{0}}d(t)+\a d_{0} f(t)\dot{d}(t),
\label{gdot}
\\
\dot{d}(t)&=&2d_{0}\w_{0}g(t),
\label{ddot}
\eea
\end{subequations}
where we introduced $\w_{0}=\e_{1}-\e_{2}$. Taking the time derivative of 
Eq.~(\ref{ddot}) and using Eq.~(\ref{gdot}) we find 
\be
\ddot{d}(t)+\w_{0}^{2}d(t)-2\a d_{0}^{2}\w_{0}f(t)\dot{d}(t)=0.
\label{step2}
\ee
With the help of Eq.~(\ref{ddot}) we rewrite Eq.~(\ref{hdot}) as 
$\dot{f}(t)=-\frac{\a}{\w_{0}}\dot{d}^{2}(t)$. Therefore $f(t)$ is a 
monotonically decreasing function of time and, by definition, 
it is bounded between $-1$ and $1$. This implies that 
$\lim_{t\ra\iif}f(t)=f_{\iif}\in(-1,1)$. If $f_{\iif}$ were
positive then the long-time solution of Eq.~(\ref{step2})  would 
be an oscillatory function with an exponentially increasing 
amplitude, in contradiction with the fact that $d(t)\in 
(d_{0},-d_{0})$. We conclude that the limiting value $f_{\iif}\in 
(-1,0)$ independently of 
the initial condition. Consequently, for large $t$ the function $d(t)$ oscillates 
at frequency $\w_{0}$ with an amplitude decaying as $e^{-\g 
t}$, where $\g=2\a d_{0}^{2}\w_{0}|f_{\iif}|$. Our analysis does 
explain the physical origin of the damping as well as the dependence 
of $\g$ on the thickness and density of the gas. In fact, 
$\g\propto\a\propto\ell_{\rm eff} n^{(\rm at)}$;
therefore the larger the thickness and/or the density is and 
the faster the amplitude of the dipole oscillations decays.

\section{Pump-induced ionization}
\label{opensec}

In this Section we extend the NEGF+GKBA formalism to deal with 
ionization processes induced by the pump. For this purpose it is
convenient to work with the HF orbitals. Let 
$\r^{\rm eq}$ be the equilibrium density matrix and 
\be
h^{\rm eq}_{{\rm HF},ij}=h_{ij}+\sum_{mn}w_{imnj}
\r^{\rm eq}_{nm}
\ee
the equilibrium HF Hamiltonian in the original 
basis. The HF orbitals 
$\q_{\m}(\blr)=\sum_{i}a_{i}^{\m}\vf_{i}(\blr)$ diagonalize both $h^{\rm 
eq}_{{\rm HF}}$ and $\r^{\rm eq}$, and are orthonormal. 
To distinguish the HF basis from the original basis we use 
greek letters to label the former. 
We have 
\be
h^{\rm eq}_{{\rm HF},\m\n}=\d_{\m\n}\e_{\m},
\ee
and $\r^{\rm 
eq}_{\m\n}=\d_{\m\n}n_{\m}$ where $n_{\m}=1$ if $\e_{\m}<\e_{\rm F}<0$ and 
$n_{\m}=0$ otherwise, $\e_{\rm F}$ being the Fermi energy. 

In finite systems like atoms and molecules the HF orbitals with $\e_{\m}>0$ 
are states in the continuum. These are the states  that get occupied by the 
photoelectron in a ionization process.
We assume that the Coulomb interaction between 
photoelectrons and bound electrons is negligible and set to zero 
the two-electron integrals $v_{\m\n\a\b}$ with at least one of the four 
indices in the continuum (this amounts to neglect Auger transitions). 
Then the self-energy has
nonvanishing matrix elements only between 
bound states. In Appendix \ref{derivation} we prove that the equation 
of motion for the density matrix with both indices in the bound sector 
reads
\be
-i\frac{d}{dt}\r(t)+\left[h_{\rm 
HF}(t),\r(t)\right]=i\left[I(t)+I_{\rm ion}(t)\right]-{\rm H.c.}.
\label{eomrhobb}
\ee
In Eq.~(\ref{eomrhobb}) every matrix has indices running over the bound states. 
The integral $I_{\rm ion}$ accounts for the pump-induced ionization 
(since $\Tr[I_{\rm ion}+{\rm H.c.}]\neq 0$ the number of bound 
electrons is 
not conserved)
and it is calculated like in Eq.~(\ref{collint}) except that 
the correlation self-energy is replaced by the ionization 
self-energy  $\S_{\rm ion}$. The latter has a vanishing 
lesser part and a greater part given by (see Appendix \ref{derivation})
\be
\S_{\rm 
ion}^{>}(t,t')=\sum_{ij}\callE_{i}(t)\s^{ij}(t-t')\callE_{j}(t').
\label{selfion>}
\ee
Here $\callE_{i}$ is the $i$-th component of the electric 
field $\bcallE=(\callE_{x},\callE_{y},\callE_{z})$ and the tensor 
\be
\s^{ij}_{\m\n}(t-t')=-i\sum_{\a\in 
c}d_{i,\m\a}\,e^{-i\e_{\a}(t-t')}\,d_{j,\a\n},
\label{cselfion>}
\ee
where $d_{i}$ is the $i$-th component of the vector of matrices
$\bld=(d_{x},d_{y},d_{z})$ and the sum over $\a$ runs in the continuum. In Fourier space
\bea
\tilde{\s}^{ij}_{\m\n}(\w)\!\!&=&\!\!-2\p i\sum_{\a\in c}d_{i,\m\a}\,\d(\w-\e_{\a})\,d_{j,\a\n}
\nn\\
\!\!&\approx&\!\!
2i\sum_{\a\in c}
d_{i,\m\a} \Im\left[\frac{1}{\w-\e_{\a}+i\eta}\right]d_{j,\a\n},
\label{sigmaion}
\eea
where $\eta$ is a positive constant of the order of the level spacing 
of the continuum states. Typically the ionization is caused by the 
action of a pump pulse with a Fourier transform peaked around some 
frequency $\w_{P}$ (larger than the ionization energy of the system). Therefore, 
$\S^{>}_{\rm ion}$ is dominated by those terms in $\s(t-t')$ that 
oscillate at frequency $\e_{\a}\simeq \w_{P}$. By virtue of this 
observation
we implement a time-local approximation
\be
\tilde{\s}^{ij}_{\m\n}(\w)\approx 
\tilde{\s}^{ij}_{\m\n}(\w_{P}),
\ee
which implies 
$\s^{ij}_{\m\n}(t-t')=\tilde{\s}^{ij}_{\m\n}(\w_{P})\d(t-t')$. 
Substituting this result into Eq.~(\ref{selfion>}) yields 
\be
\S_{\rm 
ion}^{>}(t,t')=-i\d(t-t')\G_{\m\n}(t),
\label{selfion>local}
\ee
where
\be
\G_{\m\n}(t)=i\sum_{ij}\callE_{i}(t)\tilde{\s}^{ij}_{\m\n}(\w_{P})\callE_{j}(t)
\ee
is a self-adjoint positive-definite matrix for all times $t$.

The time-local approximation allows us to simplify the integral 
$I_{\rm ion}(t)$ appearing in Eq.~(\ref{eomrhobb}). Taking into 
account that $\S_{\rm ion}^{<}=0$ we have
\bea
I_{\rm ion}(t)\!\!&=&\!\!\int_{-\iif}^{t}dt'\S^{>}_{\rm ion}(t,t')G^{<}(t',t)
\nn\\
\!\!&=&\!\!\G(t)\r(t).
\eea
Inserting this result into Eq.~(\ref{eomrhobb}) we finally obtain 
\be
-i\frac{d}{dt}\r(t)+\left[h_{\rm 
HF}(t),\r(t)\right]-i\left\{\G(t),\r(t)\right\}
=iI(t)-{\rm H.c.}
\label{eomrhobb-wbla}
\ee
where the curly brackets signify an anticommutator. Equation 
(\ref{eomrhobb-wbla}) constitutes the generalization of the NEGF+GKBA 
formalism to open systems.\cite{lpuls.2014}

\section{Krypton}
\label{Krsec}

We apply the formalism of the previous Section to address a 
retardation effect observed by Goulielmakis et al.~\cite{glwsetal.2010}
in the TPA spectrum of a krypton gas. In the experiment a strong pump is shone 
on the gas, electrons from the $4p$ shell are expelled and an admixture of Kr atoms and Kr$^{n+}$ ions, 
with $n=1,2,3\ldots$, is formed.
The admixture is subsequently probed with an 
ultrafast pulse, thus inducing transitions from the $3d$ to the 
$4p$ shell. The main 
focus in Ref.~\onlinecite{glwsetal.2010} was on 
the coherent oscillations~\cite{ps.2015,rs.2009,sypl.2011} of 
the peak intensities of the Kr$^{1+}$ ion as a 
function of the pump-probe delay $\t$. However, the 
experimental TPA spectrum reveals another interesting feature as a function of 
$\t$. The absorption 
peaks of Kr$^{2+}$ develops after the absorption 
peaks of Kr$^{1+}$, implying that 
it is faster to expel one electron than two electrons. 
To reproduce this retardation effect 
theoretically a formalism for the TPA spectrum of an 
{\em evolving admixture} is needed.
The NEGF+GKBA 
approach is in principle suitable for this purpose. As we shall see, 
important qualitative aspects of the admixture are 
intimately related to the 
diagrammatic structure of the self-energy. 
 
The Kr gas is ionized by a few-cycle NIR pump 
 $\blE(\ell \,t)=(E(t),0,0)$ with 
$E(t)=E_{0}\sin^{2}(\p t/\D_{P})\sin(\w_{P}t)$ for $0<t<\D_{P}$. The 
experimental pump intensity is $\mathfrak{I}_{0}=7\times 10^{14}$ 
W/cm$^{2}$,  
corresponding to an electric field 
$E_{0}=\sqrt{2\mathfrak{I}_{0}/c\e_{0}}=7.2\times 10^{10}$ V/m, the 
duration of the pump pulse is $\D_{P}\sim 7.6$  fs and the NIR 
frequency is $\w_{P}\sim 1.65$ eV. After a time $\t$ the Kr admixture 
is probed with an extreme ultraviolet attosecond pulse 
$\ble(\ell \,t)=(e(t),0,0)$, 
with $e(t)=e_{0}\sin^{2}(\p (t-\tilde{\tau})/\D_{p})\sin(\w_{p}(t-\tilde{\tau}))$ for 
$\tilde{\tau}<t<\tilde{\tau}+\D_{p}$. Here 
$\tau=\tilde{\tau}-(\D_{P}-\D_{p})/2$ is the time-distance between the 
maxima of the pump and probe pulses.
The probe pulse has duration
$\D_{p}\sim 150$ as, it is centered at frequency $\w_{p}=80$ eV and it 
has an intensity $\mathfrak{i}_{0}\sim  10^{11}$ 
W/cm$^{2}$, which corresponds to an electric field $e_{0}=8.6\times 
10^{8}$ V/m. We discard the dressing of the probe field
and solve the equation of motion for $\r$ with $\bcallE=\ble$.

\begin{figure}[tbp]
\includegraphics[width=8.cm]{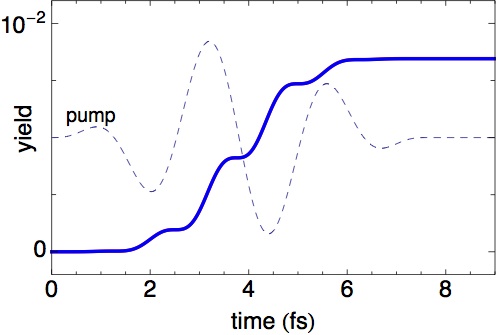}
\caption{(Color online) Transient ionization yield per spin (solid line) in HF and 2B 
(indistinguishable) and amplitude 
of the pump pulse (dashed line) in arbitrary units.}
\label{yield}
\end{figure}

The one- and two-electron 
integrals as well as  the dipole matrix elements have been calculated 
with the SMILES 
package~\cite{smiles1,smiles2} using the 
66 STO basis functions taken from Ref.~\onlinecite{bbb.1993}. As we are not interested in 
the coherent oscillations of the peak intensities we do not include the 
spin-orbit coupling responsible for the splitting of the $4p$ and 
$3d$ orbitals. Thus we should expect one main absorption peak {\em 
per ion}, corresponding to transitions from the $4p$ to the $3d$ shell.
We find eighteen HF states with energy below zero. The 
remaining HF states are used to construct the ionization self-energy 
according to Eq.~(\ref{sigmaion}). The simulations show that 
electrons are essentially removed from the $4p$ shell in agreement with the 
analysis of Ref.~\onlinecite{glwsetal.2010}. In Fig.~\ref{yield}
we display the transient ionization yield, i.e., the expelled charge 
per spin, during the action of the pump. The charge is expelled in 
pockets at a rate of twice the frequency of the laser pulse, in 
agreement with the CI calculations.~\cite{rs.2009,glwsetal.2010}
Interestingly, the HF and 2B yields are indistinguishable. 
The situation is drastically different for the TPA 
spectrum, with the 2B approximation performing much better than the 
HF one (see below).
Whether the absorption onset of Kr$^{1+}$ 
is earlier than the absorption onset of Kr$^{2+}$ is the   
central issue addressed below.

\begin{figure}[tbp]
\includegraphics[width=8.5cm]{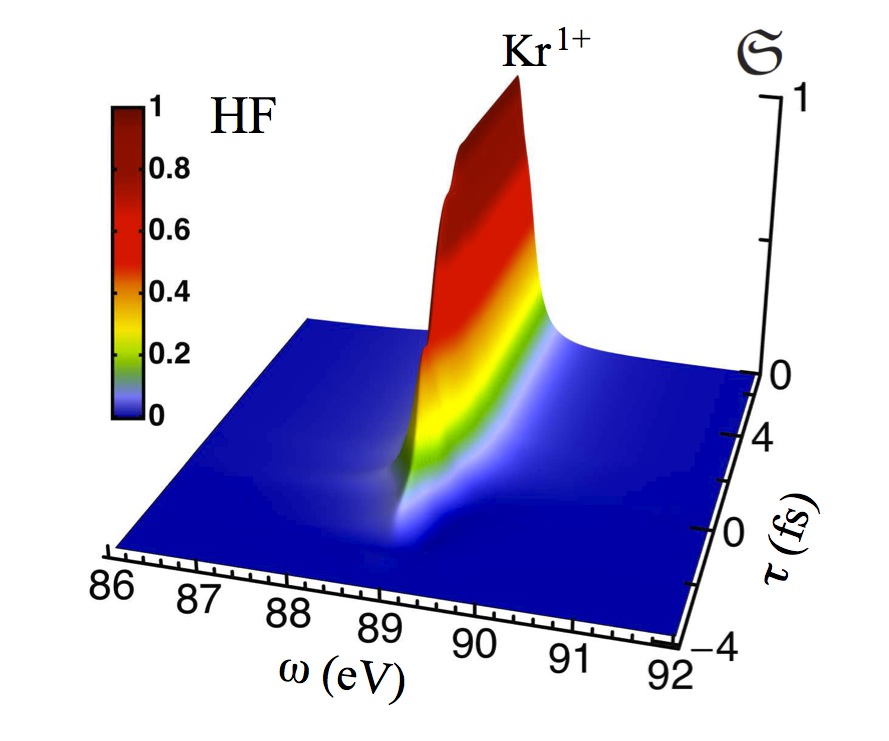}
\includegraphics[width=8.5cm]{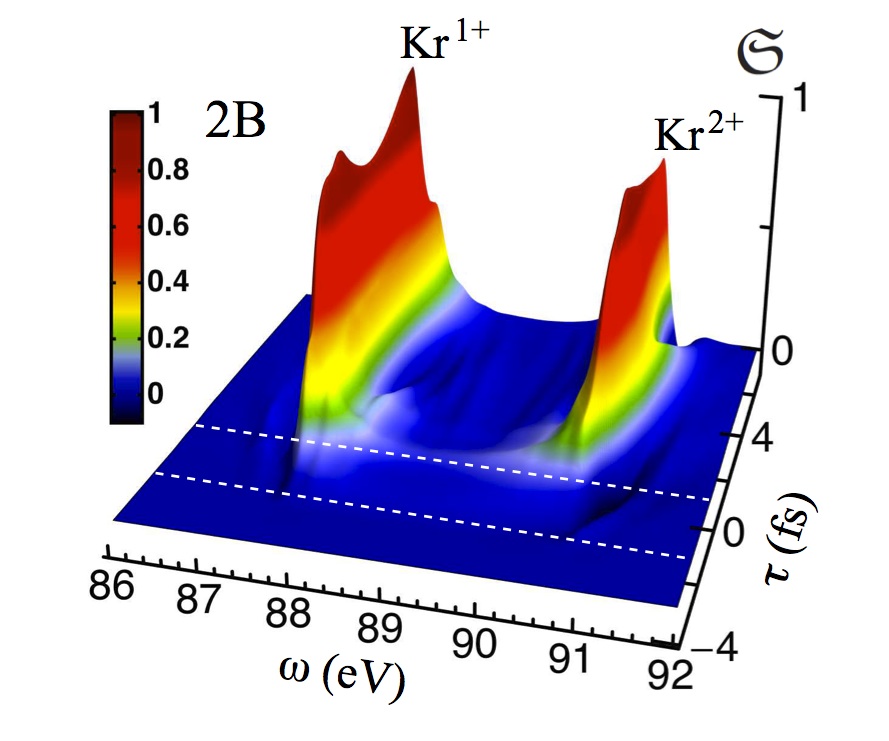}
\caption{(Color online) TPA spectrum (normalized to the maximum height)
of a krypton gas in the HF (top panel) and 2B 
(bottom panel) approximation. White (dashed) lines in the bottom panel 
are guide for the eyes to better visualize the retardation effect.
The pump and probe pulses are given in the main text.}
\label{Kr}
\end{figure}

In the upper panel of Fig.~\ref{Kr} we report the TPA spectrum of the Kr 
admixture in the HF approximation. We have propagated the system 
for $\sim~50$ fs after the action of the probe with a time step $\D 
t=0.0025$ fs (corresponding to $~$20.000 time steps) 
and then broadened the Fourier transform of $\bld_{p}(t)$
by $0.8$ eV to account for the experimental resolution.
The result is extremely disappointing. 
The HF TPA spectrum constitutes of one doublet (merged in 
Fig.~\ref{Kr} in one single 
line-shape due to the broadening) with  
a simultaneous raise of both peaks. The peaks correspond to 
transitions from the $3d$ shell to either the $4p_{x}$ orbital or the 
$4p_{y,z}$  orbitals. In fact, these transitions are nondegenerate 
in the HF approximation. As the pump 
is polarized along $x$ the 
$4p_{x}$ orbital looses more charge than the $4p_{y,z}$ orbitals, 
thereby breaking the degeneracy (albeit only slightly). 
Even more noteworthy, however, is the absence of 
spectral structures due absorption of
multiply ionized Kr atoms. The numerical simulation has been repeated with pump 
fields of different frequencies and intensities but no sign of 
other absorption peaks has been observed. The first important 
conclusion of this preliminary study is that the appearance of 
absorption peaks in multiply ionized Kr atoms is a {\em correlation effect}. 

We have then included correlation effects at the level of the  
Markovian 2B approximation but the outcome has not changed (not shown). The main 
difference between the 
HF and the Markovian 2B spectra is an overall frequency shift.
The so far accumulated numerical evidence leads us to
conjecture that any time-local approximation (no memory) to the 
self-energy is doomed to fail. We mention that the equation of motion for 
$\r$ with a time-local $\S$ has the same mathematical structure of the
Time-Dependent Density Functional Theory (TDDFT) equations at the 
level of the Adiabatic Local Density Approximation (ALDA). 
Hence, TDDFT spectra at the ALDA level would also fail in capturing 
the absorption peaks of multiply ionized atoms. The second important 
conclusion is that static correlation effects are not enough.

Dynamical correlation effects are contained in the full (nonlocal in time) 2B 
self-energy. We have solved the equations of motion for $\r$ with 
the collision integral of Eq.~(\ref{collgkba}).  
The TPA spectrum is shown in the 
lower panel of Fig.~\ref{Kr}. We clearly distinguish two structures 
corresponding to the TPA spectrum of Kr$^{1+}$ and Kr$^{2+}$.
In fact, the energy gap between the structures is consistent with the 
$\sim~3$ eV experimental gap of these two ions.\cite{notepeakpos} 
Remarkably, the high-energy structure develops $\sim~3$ fs after the 
low-energy one. This is the aforementioned retardation effect which we 
have just proved to be 
within reach of the NEGF+GKBA approach. 
Furthermore, the obtained delay is commensurate with the 5 fs delay observed 
in experiments. A non-local in time self-energy is crucial
for the appropriate description of the Kr (and probably of 
any other) evolving admixture.

Although the 2B approximation represents a noticeable improvement over 
time-local approximations
there still remains one issue to address. The experimental TPA 
spectrum contains small absorption peaks attributable to transitions in
the Kr$^{3+}$ ion. We have not been able to see these structure 
within the 2B approximation. Although we are not aware of any 
formal result relating the possibility of describing multiply ionized 
atoms to the diagrammatic structure of the self-energy we observe 
that the kernel $\d\S/\d G$  contains at most 
one particle-hole excitation in HF and two particle-hole excitations 
in 2B. It is therefore tempting to argue that in order to observe the absorption 
peaks of Kr$^{n+}$ the kernel $\d\S/\d G$ should contain at least $n$ 
particle-hole excitations,\cite{svl-book,smvl.2012} which implies that $\S$ should contain 
diagrams of order at least $n$ in the interaction. Finding a general 
solution to this problem would certainly be valuable and  
contribute to advance the understanding of many-body diagrammatic theories.

\section{Conclusions}
\label{concsec}

We have introduced a NEGF+GKBA approach to transient photoabsorption 
experiment suitable for pump fields of arbitrary strength, 
frequency and duration and for any delay between pump ad probe pulses 
(hence for delays in the overlapping regime too). The size of the 
arrays in NEGF calculations  
scales quadratically with the number of basis functions. 
The Coulomb interaction between electrons is included diagrammatically 
through the correlation self-energy and the possibility of ionization 
can be described through the ionization self-energy.

The approach has been benchmarked against the TPA 
spectrum of He reported in Ref.~\onlinecite{pbbmnl.2013}.
Helium is a weakly correlated system and all self-energy 
approximations have been shown to agree with the CI results. We have provided 
a simple yet rigorous explanation of the bending of the AT absorption 
peaks and derived a useful formula 
for fitting the experimental TPA spectra. 
We have also addressed 
the exponential damping of the probe-induced dipole and related it to 
the thickness and density of the gas. 

A more severe test for the NEGF+GKBA approach is the TPA spectrum of 
Kr reported in Ref.~\onlinecite{glwsetal.2010}. We have shown that 
for a proper description of the evolving admixture of the Kr 
ions the self-energy should have {\em memory}. 
This is not the case for the HF and Markovian 2B self-energies which 
yield the TPA spectrum of a pure Kr$^{1+}$ ion. We argue 
that the situation does not change in TDDFT with ALDA 
exchange-correlation potentials. On the contrary,
the full 2B self-energy leads to a second structure in the TPA 
spectrum that is assigned to Kr$^{2+}$ and that develops about 2-3 fs after the first, 
in fair agreement with the experiment. More theoretical and numerical 
work is needed to understand the relation between self-energy 
diagrams and the emergence of absorption peaks due to multiply 
ionized atoms. 

\section{Acknowledgments}
We thank Rafael L{\'o}pez for providing us with the SMILES 
package and Stefan Kurth for useful discussions. We 
further like to thank the CSC-IT center for science in Espoo, Finland for computing resources.
EP and GS acknowledge funding by MIUR FIRB Grant No. RBFR12SW0J. 
RvL thanks the Academy of Finland for support.

\appendix

\section{The embedded GKBA equation for $\r$}
\label{derivation}

The lesser and greater Green's functions follow from the Keldysh 
Green's function $G(z,z')$ with arguments $z$ and $z'$ on the Keldysh 
contour. In particular $G^{<}$ ($G^{>}$) is the Keldysh $G$ with the 
first (second) contour argument on the forward branch and the second 
(first) contour argument on the backward branch. The Keldysh $G$ 
satisfies the equations of motion\cite{svl-book} (in matrix form)
\bea
\big[i\frac{d}{dz}-h_{\rm HF}(z)\big]G(z,z')\!\!&=&\d(z,z')
\nn\\ &+&\!\!\int\! 
d\bar{z}\,\S(z,\bar{z})G(\bar{z},z'),
\label{eomkg}
\\
G(z,z')\big[\!-i\frac{\overleftarrow{d}}{dz'}-h_{\rm 
HF}(z')\big]\!\!&=&\d(z,z')
\nn\\ &+&\!\!\int\! 
d\bar{z}\,G(z,\bar{z})\S(\bar{z},z'),\quad\quad
\label{eomkgadj}
\eea
where the integral is over  the Keldysh 
contour.
Choosing $z$ on the backward branch and $z'$ on the forward branch, 
and applying the Langreth rules we obtain the equations of motion for 
$G^{<}$
\begin{multline}
\big[i\frac{d}{dt}-h_{\rm HF}(t)\big]G^{<}(t,t')=\int\! 
d\bar{t}\,\S^{\rm R}(t,\bar{t})G^{<}(\bar{t},t') \\
+\int\! 
d\bar{t}\,\S^{<}(t,\bar{t})G^{\rm A}(\bar{t},t').\quad\quad
\label{eom}
\end{multline}
\begin{multline}
G^{<}(t,t')\big[\!-i\frac{\overleftarrow{d}}{dt'}-h_{\rm 
HF}(t')\big]=\int\! 
d\bar{t}\,G^{\rm R}(t,\bar{t})\S^{<}(\bar{t},t') \\
+\int\! 
d\bar{t}\,G^{<}(t,\bar{t})\S^{\rm A}(\bar{t},t').
\label{adjeom}
\end{multline}
The upper indices ``R'' and ``A'' signify retarded and advanced 
functions respectively. These are defined according to
\be
F^{\rm R/A}(t,t')=\pm\th(\pm t\mp t')[F^{>}(t,t')-F^{<}(t,t')].
\ee
Subtracting Eq.~(\ref{adjeom}) from Eq.~(\ref{eom}) and setting 
$t'=t$ we find the equation of motion Eq.~(\ref{eomrho}) for the density matrix 
$\r(t)=-iG^{<}(t,t)$.

Let us work in the HF basis $\{\q_{\m}\}$ of the equilibrium system and 
write $\r_{\m\n}(t)=\d_{\m\n}n_{\m}+\d\r_{\m\n}(t)$. In the same 
basis the HF Hamiltonian in Eq.~(\ref{HFham}) reads 
\be
h_{\rm HF,\m\n}(t)=\tilde{h}_{\rm HF,\m\n}(t)+\bcallE(t)\cdot\bld_{\m\n}
\label{hHF=tilde+dip}
\ee
with
\bea
\tilde{h}_{\rm HF,\m\n}(t)&=&
h_{\m\n}+\sum_{\a\b}w_{\m\a\b\n}\r_{\b\a}(t)
\nn\\
&=&\d_{\m\n}\e_{\m}+\sum_{\a\b}w_{\m\a\b\n}\d\r_{\b\a}(t).
\label{tildehHF}
\eea
The HF states can be grouped according to their energies: if $\e_{\m}<0$ then $\q_{\m}$ is 
a bound state, otherwise $\q_{\m}$ is a continuum state. We assume 
that the two-electron integrals $v_{\m\a\b\n}$ with at least one 
index in the continuum are negligible and set them to zero.
Consequently,  $w_{\m\a\b\n}$ with at least one 
index in the continuum vanishes too.
Let us represent a matrix $\callM$ in the HF basis as
\be
\callM=\left(\begin{array}{cc}
\callM^{bb} & \callM^{bc} \\
\callM^{bc} & \callM^{cc} \\
\end{array}\right)
\label{blockmat}
\ee
where in $\callM^{bb}_{\m\n}$ both indices run over the bound states, 
in $\callM^{bc}_{\m\n}$ the first index run over the bound states and 
the second index over the continuum states, and so on.
Then, the HF Hamiltonian has following  block structure
\be
h_{\rm HF}=\left(\begin{array}{cc}
h_{\rm HF}^{bb} &  \bcallE\cdot\bld^{bc}\\
\bcallE\cdot\bld^{cb} & h_{\rm HF}^{cc} \\
\end{array}\right),
\label{blockHF}
\ee
where we took into account that $\tilde{h}_{\rm HF}^{bc}=0$, see 
Eq.~(\ref{tildehHF}). 
Similarly, we infer that the block structure of the correlation 
self-energy is
\be
\S=\left(\begin{array}{cc}
\S^{bb} & 0 \\
0 & 0 \\
\end{array}\right).
\ee

We can make use of the block structure of $h_{\rm HF}$ and $\S$ to 
simplify the equations of motion for the Keldysh $G$. In the 
bound-bound sector Eq.~(\ref{eomkg}) reads
\begin{multline}
\big[i\frac{d}{dz}-h^{bb}_{\rm HF}(z)\big]G^{bb}(z,z')
-\big[\bcallE(z)\cdot\bld^{bc}\big]G^{cb}(z,z')\\=\d(z,z')+\int\! 
d\bar{z}\,\S^{bb}(z,\bar{z})G^{bb}(\bar{z},z'),
\label{kbe-bb}
\end{multline}
whereas in the continuum-bound sector the same equation reads    
\be
\big[i\frac{d}{dz}-h^{cc}_{\rm HF}(z)\big]G^{cb}(z,z')
-\big[\bcallE(z)\cdot\bld^{cb}\big]G^{bb}(z,z')=0.
\label{kbe-cb}
\ee
We define the continuum (noninteracting) Green's function $g^{cc}$ as 
the solution of
\be
\big[i\frac{d}{dz}-h^{cc}_{\rm HF}(z)\big]g^{cc}(z,z')=\d(z,z'),
\label{gcc}
\ee
and rewrite Eq.~(\ref{kbe-cb}) in integral form
\be
G^{cb}(z,z')=\int d\bar{z}\,g^{cc}(z,\bar{z})
\big[\bcallE(\bar{z})\cdot\bld^{cb}\big]G^{bb}(\bar{z},z').
\label{gcbexpl}
\ee
Inserting Eq.~(\ref{gcbexpl}) into Eq.~(\ref{kbe-bb}) we find
\begin{multline}
\big[i\frac{d}{dz}-h^{bb}_{\rm HF}(z)\big]G^{bb}(z,z')
=\d(z,z')\\
+\int\! d\bar{z}\,\big[\S^{bb}(z,\bar{z})+\S^{bb}_{\rm ion}
(z,\bar{z})\big]G^{bb}(\bar{z},z'),
\label{kbe-bb-emb}
\end{multline}
with the ionization self-energy defined according to
\be
\S^{bb}_{\rm ion}(z,z')=\big[\bcallE(z)\cdot\bld^{bc}\big]
g^{cc}(z,z')
\big[\bcallE(z')\cdot\bld^{cb}\big].
\label{ksion}
\ee
Thus, the continuum states can be downfolded in an exact way into
an effective equation for $G^{bb}$. A similar equation can be 
derived starting from Eq.~(\ref{eomkgadj}) and reads
\begin{multline}
G^{bb}(z,z')\big[\!-i\frac{\overleftarrow{d}}{dz'}-h^{bb}_{\rm 
HF}(z')\big]\!\!=\d(z,z')
\\ \int\! 
d\bar{z}\,G^{bb}(z,\bar{z})
\big[\S^{bb}(\bar{z},z')+\S^{bb}_{\rm ion}
(\bar{z},z')\big].\quad\quad
\label{kbe-bb-emb-adj}
\end{multline}
Below we use Eqs.~(\ref{kbe-bb-emb}, \ref{kbe-bb-emb-adj}) to 
generate an equation for the density matrix in the bound-bound 
sector. To lighten the notation we  omit the upper indices ``$bb$'', so a matrix with no 
upper indices is a matrix in the  bound-bound 
sector.

Comparing Eqs.~(\ref{kbe-bb-emb}, \ref{kbe-bb-emb-adj}) with 
Eqs.~(\ref{eomkg}, \ref{eomkgadj}) we deduce that
the equations of motion for $G^{<}$ are the same as 
Eqs.~(\ref{eom}, \ref{adjeom}) except that the correlation 
self-energy is replaced by $\S+\S_{\rm ion}$. Therefore, the equation of motion 
for $\r$ is the same as Eq.~(\ref{eomrho}) except that the collision 
integral is calculated with $\S^{\lessgtr}+\S_{\rm ion}^{\lessgtr}$. 
From Eq.~(\ref{gcc}) we have
\be
g^{cc,\lessgtr}_{\m\n}(t,t')=-i\sum_{\a\in c}\callU^{cc}_{\m\a}(t)\,n^{\lessgtr}_{\a}\,
[\callU^{cc}(t')]^{\dag}_{\a\n}
\ee
where $\callU^{cc}(t)=T[e^{-i\int^{t}_{0}d t' h_{\rm HF}^{cc}(t')}]$ 
is the evolution operator in the continuum sector whereas
$n_{\a}^{<}=n_{\a}$ and $n_{\a}^{>}=1-n_{\a}$. For $\a$ in the 
continuum we have $\e_{\a}>0>\e_{\rm F}$  and hence
$n_{\a}^{<}=0$ (which implies $g^{cc,<}=0$) and $n_{\a}^{>}=1$.
The evolution operator 
takes a very simple form if we ignore the effect of the pump between 
continuum states, i.e., if we approximate $\bld^{cc}\approx 0$. In 
this case $h_{\rm HF,\m\n}^{cc}=\d_{\m\n}\e_{\m}$, see 
Eqs.~(\ref{hHF=tilde+dip}, \ref{tildehHF}), and hence
\be
g^{cc,>}_{\m\n}(t,t')=-i\d_{\m\n}e^{-i\e_{\m}(t-t')}.
\ee
From Eq.~(\ref{ksion}) it follows that the greater ionization 
self-energy is 
\be
\S_{\rm ion,\m\n}^{>}(t,t')=-i\sum_{\a}\big[\bcallE(t)\cdot\bld^{bc}_{\m\a}\big]
e^{-i\e_{\a}(t-t')}
\big[\bcallE(t')\cdot\bld^{cb}_{\a\n}\big],
\label{ksion>}
\ee
which agrees with Eqs.~(\ref{selfion>}, \ref{cselfion>}).


\begin{thebibliography}{67}
\expandafter\ifx\csname natexlab\endcsname\relax\def\natexlab#1{#1}\fi
\expandafter\ifx\csname bibnamefont\endcsname\relax
  \def\bibnamefont#1{#1}\fi
\expandafter\ifx\csname bibfnamefont\endcsname\relax
  \def\bibfnamefont#1{#1}\fi
\expandafter\ifx\csname citenamefont\endcsname\relax
  \def\citenamefont#1{#1}\fi
\expandafter\ifx\csname url\endcsname\relax
  \def\url#1{\texttt{#1}}\fi
\expandafter\ifx\csname urlprefix\endcsname\relax\def\urlprefix{URL }\fi
\providecommand{\bibinfo}[2]{#2}
\providecommand{\eprint}[2][]{\url{#2}}

\bibitem[{\citenamefont{Krausz and Ivanov}(2009)}]{ki.2009}
\bibinfo{author}{\bibfnamefont{F.}~\bibnamefont{Krausz}} \bibnamefont{and}
  \bibinfo{author}{\bibfnamefont{M.}~\bibnamefont{Ivanov}},
  \bibinfo{journal}{Rev. Mod. Phys.} \textbf{\bibinfo{volume}{81}},
  \bibinfo{pages}{163} (\bibinfo{year}{2009}).

\bibitem[{\citenamefont{Berera et~al.}(2009)\citenamefont{Berera, van
  Grondelleand, and Kennis}}]{bgk.2009}
\bibinfo{author}{\bibfnamefont{R.}~\bibnamefont{Berera}},
  \bibinfo{author}{\bibfnamefont{R.}~\bibnamefont{van Grondelleand}},
  \bibnamefont{and} \bibinfo{author}{\bibfnamefont{J.~T.~M.}
  \bibnamefont{Kennis}}, \bibinfo{journal}{Photosynth. Res.}
  \textbf{\bibinfo{volume}{101}}, \bibinfo{pages}{105} (\bibinfo{year}{2009}).

\bibitem[{\citenamefont{Sansone et~al.}(2012)\citenamefont{Sansone, Pfeifer,
  Simeonidis, and Kuleff}}]{spsk.2012}
\bibinfo{author}{\bibfnamefont{G.}~\bibnamefont{Sansone}},
  \bibinfo{author}{\bibfnamefont{T.}~\bibnamefont{Pfeifer}},
  \bibinfo{author}{\bibfnamefont{K.}~\bibnamefont{Simeonidis}},
  \bibnamefont{and} \bibinfo{author}{\bibfnamefont{A.~I.}
  \bibnamefont{Kuleff}}, \bibinfo{journal}{Chem. Phys. Chem.}
  \textbf{\bibinfo{volume}{13}}, \bibinfo{pages}{661} (\bibinfo{year}{2012}).

\bibitem[{\citenamefont{Gallmann et~al.}(2013)\citenamefont{Gallmann, Herrmann,
  Locher, Sabbar, Ludwig, Lucchini, and Keller}}]{ghlsllk.2013}
\bibinfo{author}{\bibfnamefont{L.}~\bibnamefont{Gallmann}},
  \bibinfo{author}{\bibfnamefont{J.}~\bibnamefont{Herrmann}},
  \bibinfo{author}{\bibfnamefont{R.}~\bibnamefont{Locher}},
  \bibinfo{author}{\bibfnamefont{M.}~\bibnamefont{Sabbar}},
  \bibinfo{author}{\bibfnamefont{A.}~\bibnamefont{Ludwig}},
  \bibinfo{author}{\bibfnamefont{M.}~\bibnamefont{Lucchini}}, \bibnamefont{and}
  \bibinfo{author}{\bibfnamefont{U.}~\bibnamefont{Keller}},
  \bibinfo{journal}{Mol. Phys.} \textbf{\bibinfo{volume}{111}},
  \bibinfo{pages}{2243} (\bibinfo{year}{2013}).

\bibitem[{\citenamefont{Kuleff and Cederbaum}(2014)}]{kc.2014}
\bibinfo{author}{\bibfnamefont{A.~I.} \bibnamefont{Kuleff}} \bibnamefont{and}
  \bibinfo{author}{\bibfnamefont{L.~S.} \bibnamefont{Cederbaum}},
  \bibinfo{journal}{J. Phys. B: At. Mol. Opt. Phys.}
  \textbf{\bibinfo{volume}{47}}, \bibinfo{pages}{124002}
  (\bibinfo{year}{2014}).

\bibitem[{\citenamefont{Gaarde et~al.}(2011)\citenamefont{Gaarde, Buth, Tate,
  and Schafer}}]{gbts.2011}
\bibinfo{author}{\bibfnamefont{M.~B.} \bibnamefont{Gaarde}},
  \bibinfo{author}{\bibfnamefont{C.}~\bibnamefont{Buth}},
  \bibinfo{author}{\bibfnamefont{J.~L.} \bibnamefont{Tate}}, \bibnamefont{and}
  \bibinfo{author}{\bibfnamefont{K.~J.} \bibnamefont{Schafer}},
  \bibinfo{journal}{Phys. Rev. A} \textbf{\bibinfo{volume}{83}},
  \bibinfo{pages}{013419} (\bibinfo{year}{2011}).

\bibitem[{\citenamefont{Pabst et~al.}(2011)\citenamefont{Pabst, Greenman, Ho,
  Mazziotti, and Santra}}]{pghmetal.2011}
\bibinfo{author}{\bibfnamefont{S.}~\bibnamefont{Pabst}},
  \bibinfo{author}{\bibfnamefont{L.}~\bibnamefont{Greenman}},
  \bibinfo{author}{\bibfnamefont{P.~J.} \bibnamefont{Ho}},
  \bibinfo{author}{\bibfnamefont{D.~A.} \bibnamefont{Mazziotti}},
  \bibnamefont{and} \bibinfo{author}{\bibfnamefont{R.}~\bibnamefont{Santra}},
  \bibinfo{journal}{Phys. Rev. Lett.} \textbf{\bibinfo{volume}{106}},
  \bibinfo{pages}{053003} (\bibinfo{year}{2011}).

\bibitem[{\citenamefont{Chu and Lin}(2012)}]{cl.2012}
\bibinfo{author}{\bibfnamefont{W.}~\bibnamefont{Chu}} \bibnamefont{and}
  \bibinfo{author}{\bibfnamefont{C.~D.} \bibnamefont{Lin}},
  \bibinfo{journal}{Phys. Rev. A} \textbf{\bibinfo{volume}{85}},
  \bibinfo{pages}{013409} (\bibinfo{year}{2012}).

\bibitem[{\citenamefont{Tarana and Greene}(2012)}]{tg.2012}
\bibinfo{author}{\bibfnamefont{M.}~\bibnamefont{Tarana}} \bibnamefont{and}
  \bibinfo{author}{\bibfnamefont{C.~H.} \bibnamefont{Greene}},
  \bibinfo{journal}{Phys. Rev. A} \textbf{\bibinfo{volume}{85}},
  \bibinfo{pages}{013411} (\bibinfo{year}{2012}).

\bibitem[{\citenamefont{Pabst et~al.}(2012)\citenamefont{Pabst, Sytcheva,
  Moulet, Wirth, Goulielmakis, and Santra}}]{psmwgs.2012}
\bibinfo{author}{\bibfnamefont{S.}~\bibnamefont{Pabst}},
  \bibinfo{author}{\bibfnamefont{A.}~\bibnamefont{Sytcheva}},
  \bibinfo{author}{\bibfnamefont{A.}~\bibnamefont{Moulet}},
  \bibinfo{author}{\bibfnamefont{A.}~\bibnamefont{Wirth}},
  \bibinfo{author}{\bibfnamefont{E.}~\bibnamefont{Goulielmakis}},
  \bibnamefont{and} \bibinfo{author}{\bibfnamefont{R.}~\bibnamefont{Santra}},
  \bibinfo{journal}{Phys. Rev. A} \textbf{\bibinfo{volume}{86}},
  \bibinfo{pages}{063411} (\bibinfo{year}{2012}).

\bibitem[{\citenamefont{Pfeiffer et~al.}(2013)\citenamefont{Pfeiffer, Bell,
  Beck, Mashiko, Neumark, and Leone}}]{pbbmnl.2013}
\bibinfo{author}{\bibfnamefont{A.~N.} \bibnamefont{Pfeiffer}},
  \bibinfo{author}{\bibfnamefont{M.~J.} \bibnamefont{Bell}},
  \bibinfo{author}{\bibfnamefont{A.~R.} \bibnamefont{Beck}},
  \bibinfo{author}{\bibfnamefont{H.}~\bibnamefont{Mashiko}},
  \bibinfo{author}{\bibfnamefont{D.~M.} \bibnamefont{Neumark}},
  \bibnamefont{and} \bibinfo{author}{\bibfnamefont{S.~R.} \bibnamefont{Leone}},
  \bibinfo{journal}{Phys. Rev. A} \textbf{\bibinfo{volume}{88}},
  \bibinfo{pages}{051402} (\bibinfo{year}{2013}).

\bibitem[{\citenamefont{Rohringer and Santra}(2009)}]{rs.2009}
\bibinfo{author}{\bibfnamefont{N.}~\bibnamefont{Rohringer}} \bibnamefont{and}
  \bibinfo{author}{\bibfnamefont{R.}~\bibnamefont{Santra}},
  \bibinfo{journal}{Phys. Rev. A} \textbf{\bibinfo{volume}{79}},
  \bibinfo{pages}{053402} (\bibinfo{year}{2009}).

\bibitem[{\citenamefont{Santra et~al.}(2011)\citenamefont{Santra, Yakovlev,
  Pfeifer, and Loh}}]{sypl.2011}
\bibinfo{author}{\bibfnamefont{R.}~\bibnamefont{Santra}},
  \bibinfo{author}{\bibfnamefont{V.~S.} \bibnamefont{Yakovlev}},
  \bibinfo{author}{\bibfnamefont{T.}~\bibnamefont{Pfeifer}}, \bibnamefont{and}
  \bibinfo{author}{\bibfnamefont{Z.}~\bibnamefont{Loh}},
  \bibinfo{journal}{Phys. Rev. A} \textbf{\bibinfo{volume}{83}},
  \bibinfo{pages}{033405} (\bibinfo{year}{2011}).

\bibitem[{\citenamefont{Baggesen et~al.}(2012)\citenamefont{Baggesen, Lindroth,
  and Madsen}}]{blm.2012}
\bibinfo{author}{\bibfnamefont{J.~C.} \bibnamefont{Baggesen}},
  \bibinfo{author}{\bibfnamefont{E.}~\bibnamefont{Lindroth}}, \bibnamefont{and}
  \bibinfo{author}{\bibfnamefont{L.~B.} \bibnamefont{Madsen}},
  \bibinfo{journal}{Phys. Rev. A} \textbf{\bibinfo{volume}{85}},
  \bibinfo{pages}{013415} (\bibinfo{year}{2012}).

\bibitem[{\citenamefont{Perfetto and Stefanucci}(2015)}]{ps.2015}
\bibinfo{author}{\bibfnamefont{E.}~\bibnamefont{Perfetto}} \bibnamefont{and}
  \bibinfo{author}{\bibfnamefont{G.}~\bibnamefont{Stefanucci}},
  \bibinfo{journal}{Phys. Rev. A} \textbf{\bibinfo{volume}{91}},
  \bibinfo{pages}{033416} (\bibinfo{year}{2015}).

\bibitem[{\citenamefont{De\;Giovannini
  et~al.}(2013)\citenamefont{De\;Giovannini, Brunetto, Castro, Walkenhorst, and
  Rubio}}]{dbcwr.2013}
\bibinfo{author}{\bibfnamefont{U.}~\bibnamefont{De\;Giovannini}},
  \bibinfo{author}{\bibfnamefont{G.}~\bibnamefont{Brunetto}},
  \bibinfo{author}{\bibfnamefont{A.}~\bibnamefont{Castro}},
  \bibinfo{author}{\bibfnamefont{J.}~\bibnamefont{Walkenhorst}},
  \bibnamefont{and} \bibinfo{author}{\bibfnamefont{A.}~\bibnamefont{Rubio}},
  \bibinfo{journal}{Chem. Phys. Chem} \textbf{\bibinfo{volume}{14}},
  \bibinfo{pages}{1363} (\bibinfo{year}{2013}).

\bibitem[{\citenamefont{Neidel et~al.}(2013)\citenamefont{Neidel, Klei, Yang,
  Rouz\'ee, Vrakking, Kl\"under, Miranda, Arnold, Fordell, L'Huillier
  et~al.}}]{nkyetal.2013}
\bibinfo{author}{\bibfnamefont{C.}~\bibnamefont{Neidel}},
  \bibinfo{author}{\bibfnamefont{J.}~\bibnamefont{Klei}},
  \bibinfo{author}{\bibfnamefont{C.-H.} \bibnamefont{Yang}},
  \bibinfo{author}{\bibfnamefont{A.}~\bibnamefont{Rouz\'ee}},
  \bibinfo{author}{\bibfnamefont{M.~J.~J.} \bibnamefont{Vrakking}},
  \bibinfo{author}{\bibfnamefont{K.}~\bibnamefont{Kl\"under}},
  \bibinfo{author}{\bibfnamefont{M.}~\bibnamefont{Miranda}},
  \bibinfo{author}{\bibfnamefont{C.~L.} \bibnamefont{Arnold}},
  \bibinfo{author}{\bibfnamefont{T.}~\bibnamefont{Fordell}},
  \bibinfo{author}{\bibfnamefont{A.}~\bibnamefont{L'Huillier}},
  \bibnamefont{et~al.}, \bibinfo{journal}{Phys. Rev. Lett.}
  \textbf{\bibinfo{volume}{111}}, \bibinfo{pages}{033001}
  (\bibinfo{year}{2013}).

\bibitem[{\citenamefont{Falke et~al.}(2014)\citenamefont{Falke, Rozzi, Brida,
  Maiuri, Amato, Sommer, Sio, Rubio, Cerullo, Molinari et~al.}}]{frbmetal.2014}
\bibinfo{author}{\bibfnamefont{S.~M.} \bibnamefont{Falke}},
  \bibinfo{author}{\bibfnamefont{C.~A.} \bibnamefont{Rozzi}},
  \bibinfo{author}{\bibfnamefont{D.}~\bibnamefont{Brida}},
  \bibinfo{author}{\bibfnamefont{M.}~\bibnamefont{Maiuri}},
  \bibinfo{author}{\bibfnamefont{M.}~\bibnamefont{Amato}},
  \bibinfo{author}{\bibfnamefont{E.}~\bibnamefont{Sommer}},
  \bibinfo{author}{\bibfnamefont{A.~D.} \bibnamefont{Sio}},
  \bibinfo{author}{\bibfnamefont{A.}~\bibnamefont{Rubio}},
  \bibinfo{author}{\bibfnamefont{G.}~\bibnamefont{Cerullo}},
  \bibinfo{author}{\bibfnamefont{E.}~\bibnamefont{Molinari}},
  \bibnamefont{et~al.}, \bibinfo{journal}{Science}
  \textbf{\bibinfo{volume}{344}}, \bibinfo{pages}{1001} (\bibinfo{year}{2014}).

\bibitem[{\citenamefont{Rozzi et~al.}(2013)\citenamefont{Rozzi, Falke,
  Spallanzani, Rubio, Molinari, Brida, Maiuri, Cerullo, Schramm, Christoffers
  et~al.}}]{rfsrmetal.2013}
\bibinfo{author}{\bibfnamefont{C.~A.} \bibnamefont{Rozzi}},
  \bibinfo{author}{\bibfnamefont{S.~M.} \bibnamefont{Falke}},
  \bibinfo{author}{\bibfnamefont{N.}~\bibnamefont{Spallanzani}},
  \bibinfo{author}{\bibfnamefont{A.}~\bibnamefont{Rubio}},
  \bibinfo{author}{\bibfnamefont{E.}~\bibnamefont{Molinari}},
  \bibinfo{author}{\bibfnamefont{D.}~\bibnamefont{Brida}},
  \bibinfo{author}{\bibfnamefont{M.}~\bibnamefont{Maiuri}},
  \bibinfo{author}{\bibfnamefont{G.}~\bibnamefont{Cerullo}},
  \bibinfo{author}{\bibfnamefont{H.}~\bibnamefont{Schramm}},
  \bibinfo{author}{\bibfnamefont{J.}~\bibnamefont{Christoffers}},
  \bibnamefont{et~al.}, \bibinfo{journal}{Nature Comm.}
  \textbf{\bibinfo{volume}{4}}, \bibinfo{pages}{1602} (\bibinfo{year}{2013}).

\bibitem[{\citenamefont{Maitra et~al.}(2004)\citenamefont{Maitra, Zhang, Cave,
  and Burke}}]{mzcb.2004}
\bibinfo{author}{\bibfnamefont{N.~T.} \bibnamefont{Maitra}},
  \bibinfo{author}{\bibfnamefont{F.}~\bibnamefont{Zhang}},
  \bibinfo{author}{\bibfnamefont{R.~J.} \bibnamefont{Cave}}, \bibnamefont{and}
  \bibinfo{author}{\bibfnamefont{K.}~\bibnamefont{Burke}}, \bibinfo{journal}{J.
  Chem. Phys.} \textbf{\bibinfo{volume}{120}}, \bibinfo{pages}{5932}
  (\bibinfo{year}{2004}).

\bibitem[{\citenamefont{K\"ummel and Kronik}(2008)}]{kk.2008}
\bibinfo{author}{\bibfnamefont{S.}~\bibnamefont{K\"ummel}} \bibnamefont{and}
  \bibinfo{author}{\bibfnamefont{L.}~\bibnamefont{Kronik}},
  \bibinfo{journal}{Rev. Mod. Phys.} \textbf{\bibinfo{volume}{80}},
  \bibinfo{pages}{3} (\bibinfo{year}{2008}).

\bibitem[{\citenamefont{Gritsenko and Baerends}(2004)}]{ngb.2006}
\bibinfo{author}{\bibfnamefont{O.}~\bibnamefont{Gritsenko}} \bibnamefont{and}
  \bibinfo{author}{\bibfnamefont{E.~J.} \bibnamefont{Baerends}},
  \bibinfo{journal}{J. Chem. Phys.} \textbf{\bibinfo{volume}{121}},
  \bibinfo{pages}{655} (\bibinfo{year}{2004}).

\bibitem[{\citenamefont{Maitra}(2005)}]{m.2005}
\bibinfo{author}{\bibfnamefont{N.~T.} \bibnamefont{Maitra}},
  \bibinfo{journal}{J. Chem. Phys.} \textbf{\bibinfo{volume}{122}},
  \bibinfo{pages}{234104} (\bibinfo{year}{2005}).

\bibitem[{\citenamefont{Maitra and Tempel}(2006)}]{mt.2006}
\bibinfo{author}{\bibfnamefont{N.~T.} \bibnamefont{Maitra}} \bibnamefont{and}
  \bibinfo{author}{\bibfnamefont{D.~G.} \bibnamefont{Tempel}},
  \bibinfo{journal}{J. Chem. Phys.} \textbf{\bibinfo{volume}{125}},
  \bibinfo{pages}{184111} (\bibinfo{year}{2006}).

\bibitem[{\citenamefont{Neaton et~al.}(2006)\citenamefont{Neaton, Hybertsen,
  and Louie}}]{nhl.2006}
\bibinfo{author}{\bibfnamefont{J.~B.} \bibnamefont{Neaton}},
  \bibinfo{author}{\bibfnamefont{M.~S.} \bibnamefont{Hybertsen}},
  \bibnamefont{and} \bibinfo{author}{\bibfnamefont{S.~G.} \bibnamefont{Louie}},
  \bibinfo{journal}{Phys. Rev. Lett.} \textbf{\bibinfo{volume}{97}},
  \bibinfo{pages}{216405} (\bibinfo{year}{2006}).

\bibitem[{\citenamefont{Souza et~al.}(2013)\citenamefont{Souza, Rungger,
  Pemmaraju, Schwingenschloegl, and Sanvito}}]{srpss.2013}
\bibinfo{author}{\bibfnamefont{A.~M.} \bibnamefont{Souza}},
  \bibinfo{author}{\bibfnamefont{I.}~\bibnamefont{Rungger}},
  \bibinfo{author}{\bibfnamefont{C.~D.} \bibnamefont{Pemmaraju}},
  \bibinfo{author}{\bibfnamefont{U.}~\bibnamefont{Schwingenschloegl}},
  \bibnamefont{and} \bibinfo{author}{\bibfnamefont{S.}~\bibnamefont{Sanvito}},
  \bibinfo{journal}{Phys. Rev. B} \textbf{\bibinfo{volume}{88}},
  \bibinfo{pages}{165112} (\bibinfo{year}{2013}).

\bibitem[{\citenamefont{Stefanucci and Kurth}(2011)}]{sk.2011}
\bibinfo{author}{\bibfnamefont{G.}~\bibnamefont{Stefanucci}} \bibnamefont{and}
  \bibinfo{author}{\bibfnamefont{S.}~\bibnamefont{Kurth}},
  \bibinfo{journal}{Phys. Rev. Lett.} \textbf{\bibinfo{volume}{107}},
  \bibinfo{pages}{216401} (\bibinfo{year}{2011}).

\bibitem[{\citenamefont{Kurth and Stefanucci}(2013)}]{ks.2013}
\bibinfo{author}{\bibfnamefont{S.}~\bibnamefont{Kurth}} \bibnamefont{and}
  \bibinfo{author}{\bibfnamefont{G.}~\bibnamefont{Stefanucci}},
  \bibinfo{journal}{Phys. Rev. Lett.} \textbf{\bibinfo{volume}{111}},
  \bibinfo{pages}{030601} (\bibinfo{year}{2013}).

\bibitem[{kb-()}]{kb-book}
\bibinfo{note}{L.~P. Kadanoff and G. Baym, {\em Quantum Statistical Mechanics}
  (W. A. Benjamin, Inc. New York, 1962)}.

\bibitem[{hj-()}]{hj-book}
\bibinfo{note}{H. Haug and A.-P. Jauho, {\em Quantum Kinetics in Transport and
  Optics of Semiconductors} (Springer, Berlin, 2007)}.

\bibitem[{svl()}]{svl-book}
\bibinfo{note}{G. Stefanucci and R. van Leeuwen, {\em Nonequilibrium Many-Body
  Theory of Quantum Systems: A Modern Introduction} (Cambridge University
  Press, Cambridge, 2013)}.

\bibitem[{bb-()}]{bb-book}
\bibinfo{note}{K. Balzer, and M. Bonitz, {\em Nonequilibrium Green's Functions
  approach to Inhomogeneous Systems}, Lect. Notes Phys. {\bf 867} (2013)}.

\bibitem[{\citenamefont{Perfetto et~al.}(2015)\citenamefont{Perfetto, Sangalli,
  Marini, and Stefanucci}}]{psms.2015}
\bibinfo{author}{\bibfnamefont{E.}~\bibnamefont{Perfetto}},
  \bibinfo{author}{\bibfnamefont{D.}~\bibnamefont{Sangalli}},
  \bibinfo{author}{\bibfnamefont{A.}~\bibnamefont{Marini}}, \bibnamefont{and}
  \bibinfo{author}{\bibfnamefont{G.}~\bibnamefont{Stefanucci}},
  \bibinfo{journal}{arXiv:1507.01786}  (\bibinfo{year}{2015}).

\bibitem[{\citenamefont{Goulielmakis et~al.}(2010)\citenamefont{Goulielmakis,
  Loh, Wirth, Santra, Rohringer, Yakovlev, Zherebtsov, Pfeifer, Azzeer, Kling
  et~al.}}]{glwsetal.2010}
\bibinfo{author}{\bibfnamefont{E.}~\bibnamefont{Goulielmakis}},
  \bibinfo{author}{\bibfnamefont{Z.}~\bibnamefont{Loh}},
  \bibinfo{author}{\bibfnamefont{A.}~\bibnamefont{Wirth}},
  \bibinfo{author}{\bibfnamefont{R.}~\bibnamefont{Santra}},
  \bibinfo{author}{\bibfnamefont{N.}~\bibnamefont{Rohringer}},
  \bibinfo{author}{\bibfnamefont{V.~S.} \bibnamefont{Yakovlev}},
  \bibinfo{author}{\bibfnamefont{S.}~\bibnamefont{Zherebtsov}},
  \bibinfo{author}{\bibfnamefont{T.}~\bibnamefont{Pfeifer}},
  \bibinfo{author}{\bibfnamefont{A.~M.} \bibnamefont{Azzeer}},
  \bibinfo{author}{\bibfnamefont{M.~F.} \bibnamefont{Kling}},
  \bibnamefont{et~al.}, \bibinfo{journal}{Nature}
  \textbf{\bibinfo{volume}{466}}, \bibinfo{pages}{739} (\bibinfo{year}{2010}).

\bibitem[{\citenamefont{Lipavsk\'y et~al.}(1986)\citenamefont{Lipavsk\'y,
  $\check{\rm S}$pi$\check{\rm c}$ka, and Velick\'y}}]{lsv.1986}
\bibinfo{author}{\bibfnamefont{P.}~\bibnamefont{Lipavsk\'y}},
  \bibinfo{author}{\bibfnamefont{V.}~\bibnamefont{$\check{\rm S}$pi$\check{\rm
  c}$ka}}, \bibnamefont{and}
  \bibinfo{author}{\bibfnamefont{B.}~\bibnamefont{Velick\'y}},
  \bibinfo{journal}{Phys. Rev. B} \textbf{\bibinfo{volume}{34}},
  \bibinfo{pages}{6933} (\bibinfo{year}{1986}).

\bibitem[{\citenamefont{Bonitz et~al.}(1996)\citenamefont{Bonitz, Kremp, Scott,
  Binder, Kraeft, and K\"ohler}}]{bksbkk.1996}
\bibinfo{author}{\bibfnamefont{M.}~\bibnamefont{Bonitz}},
  \bibinfo{author}{\bibfnamefont{D.}~\bibnamefont{Kremp}},
  \bibinfo{author}{\bibfnamefont{D.~C.} \bibnamefont{Scott}},
  \bibinfo{author}{\bibfnamefont{R.}~\bibnamefont{Binder}},
  \bibinfo{author}{\bibfnamefont{W.~D.} \bibnamefont{Kraeft}},
  \bibnamefont{and} \bibinfo{author}{\bibfnamefont{H.~S.}
  \bibnamefont{K\"ohler}}, \bibinfo{journal}{J. Phys.: Condens. Matter}
  \textbf{\bibinfo{volume}{8}}, \bibinfo{pages}{6057} (\bibinfo{year}{1996}).

\bibitem[{\citenamefont{Kwong et~al.}(1998)\citenamefont{Kwong, Bonitz, Binder,
  , and K\"ohler}}]{kbbk.1998}
\bibinfo{author}{\bibfnamefont{N.~H.} \bibnamefont{Kwong}},
  \bibinfo{author}{\bibfnamefont{M.}~\bibnamefont{Bonitz}},
  \bibinfo{author}{\bibfnamefont{R.}~\bibnamefont{Binder}}, , \bibnamefont{and}
  \bibinfo{author}{\bibfnamefont{H.~S.} \bibnamefont{K\"ohler}},
  \bibinfo{journal}{Phys. Status Solidi B} \textbf{\bibinfo{volume}{206}},
  \bibinfo{pages}{197} (\bibinfo{year}{1998}).

\bibitem[{\citenamefont{Haug}(1992)}]{h.1992}
\bibinfo{author}{\bibfnamefont{H.}~\bibnamefont{Haug}}, \bibinfo{journal}{Phys.
  Status Solidi B} \textbf{\bibinfo{volume}{173}}, \bibinfo{pages}{139}
  (\bibinfo{year}{1992}).

\bibitem[{\citenamefont{Binder et~al.}(1997)\citenamefont{Binder, K\"ohler,
  Bonitz, and Kwong}}]{bkbk.1997}
\bibinfo{author}{\bibfnamefont{R.}~\bibnamefont{Binder}},
  \bibinfo{author}{\bibfnamefont{H.~S.} \bibnamefont{K\"ohler}},
  \bibinfo{author}{\bibfnamefont{M.}~\bibnamefont{Bonitz}}, \bibnamefont{and}
  \bibinfo{author}{\bibfnamefont{N.}~\bibnamefont{Kwong}},
  \bibinfo{journal}{Phys. Rev. B} \textbf{\bibinfo{volume}{55}},
  \bibinfo{pages}{5110} (\bibinfo{year}{1997}).

\bibitem[{\citenamefont{Bonitz et~al.}(1999)\citenamefont{Bonitz, Semkat, and
  Haug}}]{bsh.1999}
\bibinfo{author}{\bibfnamefont{M.}~\bibnamefont{Bonitz}},
  \bibinfo{author}{\bibfnamefont{D.}~\bibnamefont{Semkat}}, \bibnamefont{and}
  \bibinfo{author}{\bibfnamefont{H.}~\bibnamefont{Haug}},
  \bibinfo{journal}{Eur. Phys. J. B} \textbf{\bibinfo{volume}{9}},
  \bibinfo{pages}{209} (\bibinfo{year}{1999}).

\bibitem[{\citenamefont{Gartner et~al.}(1999)\citenamefont{Gartner, B\'anyai,
  and Haug}}]{gbh.1999}
\bibinfo{author}{\bibfnamefont{P.}~\bibnamefont{Gartner}},
  \bibinfo{author}{\bibfnamefont{L.}~\bibnamefont{B\'anyai}}, \bibnamefont{and}
  \bibinfo{author}{\bibfnamefont{H.}~\bibnamefont{Haug}},
  \bibinfo{journal}{Phys. Rev. B} \textbf{\bibinfo{volume}{60}},
  \bibinfo{pages}{14234} (\bibinfo{year}{1999}).

\bibitem[{\citenamefont{Vu and Haug}(2000)}]{vh.2000}
\bibinfo{author}{\bibfnamefont{Q.~T.} \bibnamefont{Vu}} \bibnamefont{and}
  \bibinfo{author}{\bibfnamefont{H.}~\bibnamefont{Haug}},
  \bibinfo{journal}{Phys. Rev. B} \textbf{\bibinfo{volume}{62}},
  \bibinfo{pages}{7179} (\bibinfo{year}{2000}).

\bibitem[{\citenamefont{Marini}(2013)}]{m.2013}
\bibinfo{author}{\bibfnamefont{A.}~\bibnamefont{Marini}}, \bibinfo{journal}{J.
  Phys: Conf. Proc.} \textbf{\bibinfo{volume}{427}}, \bibinfo{pages}{012003}
  (\bibinfo{year}{2013}).

\bibitem[{\citenamefont{Sangalli and Marini}(2015)}]{sm.2015}
\bibinfo{author}{\bibfnamefont{D.}~\bibnamefont{Sangalli}} \bibnamefont{and}
  \bibinfo{author}{\bibfnamefont{A.}~\bibnamefont{Marini}},
  \bibinfo{journal}{arXiv:1409.1706}  (\bibinfo{year}{2015}).

\bibitem[{\citenamefont{Balzer et~al.}(2013)\citenamefont{Balzer, Hermanns, and
  Bonitz}}]{bhb.2013}
\bibinfo{author}{\bibfnamefont{K.}~\bibnamefont{Balzer}},
  \bibinfo{author}{\bibfnamefont{S.}~\bibnamefont{Hermanns}}, \bibnamefont{and}
  \bibinfo{author}{\bibfnamefont{M.}~\bibnamefont{Bonitz}},
  \bibinfo{journal}{J. Phys.: Conf. Ser.} \textbf{\bibinfo{volume}{427}},
  \bibinfo{pages}{012006} (\bibinfo{year}{2013}).

\bibitem[{\citenamefont{Bonitz et~al.}(2013)\citenamefont{Bonitz, Balzer, and
  Hermanns}}]{bhb2.2013}
\bibinfo{author}{\bibfnamefont{M.}~\bibnamefont{Bonitz}},
  \bibinfo{author}{\bibfnamefont{K.}~\bibnamefont{Balzer}}, \bibnamefont{and}
  \bibinfo{author}{\bibfnamefont{S.}~\bibnamefont{Hermanns}},
  \bibinfo{journal}{Contrib. Plasma Phys.} \textbf{\bibinfo{volume}{53}},
  \bibinfo{pages}{778} (\bibinfo{year}{2013}).

\bibitem[{\citenamefont{Hermanns et~al.}(2014)\citenamefont{Hermanns,
  Schl\"unzen, and Bonitz}}]{hsb.2014}
\bibinfo{author}{\bibfnamefont{S.}~\bibnamefont{Hermanns}},
  \bibinfo{author}{\bibfnamefont{N.}~\bibnamefont{Schl\"unzen}},
  \bibnamefont{and} \bibinfo{author}{\bibfnamefont{M.}~\bibnamefont{Bonitz}},
  \bibinfo{journal}{Phys. Rev. B} \textbf{\bibinfo{volume}{90}},
  \bibinfo{pages}{125111} (\bibinfo{year}{2014}).

\bibitem[{\citenamefont{Latini et~al.}(2014)\citenamefont{Latini, , Perfetto,
  Uimonen, van Leeuwen, and Stefanucci}}]{lpuls.2014}
\bibinfo{author}{\bibfnamefont{S.}~\bibnamefont{Latini}}, ,
  \bibinfo{author}{\bibfnamefont{E.}~\bibnamefont{Perfetto}},
  \bibinfo{author}{\bibfnamefont{A.-M.} \bibnamefont{Uimonen}},
  \bibinfo{author}{\bibfnamefont{R.}~\bibnamefont{van Leeuwen}},
  \bibnamefont{and}
  \bibinfo{author}{\bibfnamefont{G.}~\bibnamefont{Stefanucci}},
  \bibinfo{journal}{Phys. Rev. B} \textbf{\bibinfo{volume}{89}},
  \bibinfo{pages}{075306} (\bibinfo{year}{2014}).

\bibitem[{\citenamefont{Dahlen and van Leeuwen}(2007)}]{dl.2007}
\bibinfo{author}{\bibfnamefont{N.~E.} \bibnamefont{Dahlen}} \bibnamefont{and}
  \bibinfo{author}{\bibfnamefont{R.}~\bibnamefont{van Leeuwen}},
  \bibinfo{journal}{Phys. Rev. Lett.} \textbf{\bibinfo{volume}{98}},
  \bibinfo{pages}{153004} (\bibinfo{year}{2007}).

\bibitem[{\citenamefont{My\"oh\"anen et~al.}(2008)\citenamefont{My\"oh\"anen,
  Stan, Stefanucci, and van Leeuwen}}]{mssvl.2008}
\bibinfo{author}{\bibfnamefont{P.}~\bibnamefont{My\"oh\"anen}},
  \bibinfo{author}{\bibfnamefont{A.}~\bibnamefont{Stan}},
  \bibinfo{author}{\bibfnamefont{G.}~\bibnamefont{Stefanucci}},
  \bibnamefont{and} \bibinfo{author}{\bibfnamefont{R.}~\bibnamefont{van
  Leeuwen}}, \bibinfo{journal}{EPL} \textbf{\bibinfo{volume}{84}},
  \bibinfo{pages}{67001} (\bibinfo{year}{2008}).

\bibitem[{\citenamefont{My\"oh\"anen et~al.}(2009)\citenamefont{My\"oh\"anen,
  Stan, Stefanucci, and van Leeuwen}}]{mssvl.2009}
\bibinfo{author}{\bibfnamefont{P.}~\bibnamefont{My\"oh\"anen}},
  \bibinfo{author}{\bibfnamefont{A.}~\bibnamefont{Stan}},
  \bibinfo{author}{\bibfnamefont{G.}~\bibnamefont{Stefanucci}},
  \bibnamefont{and} \bibinfo{author}{\bibfnamefont{R.}~\bibnamefont{van
  Leeuwen}}, \bibinfo{journal}{Phys. Rev. B} \textbf{\bibinfo{volume}{80}},
  \bibinfo{pages}{115107} (\bibinfo{year}{2009}).

\bibitem[{\citenamefont{Balzer et~al.}(2009)\citenamefont{Balzer, Bonitz, van
  Leeuwen, Stan, and Dahlen}}]{bbvlsd.2009}
\bibinfo{author}{\bibfnamefont{K.}~\bibnamefont{Balzer}},
  \bibinfo{author}{\bibfnamefont{M.}~\bibnamefont{Bonitz}},
  \bibinfo{author}{\bibfnamefont{R.}~\bibnamefont{van Leeuwen}},
  \bibinfo{author}{\bibfnamefont{A.}~\bibnamefont{Stan}}, \bibnamefont{and}
  \bibinfo{author}{\bibfnamefont{N.~E.} \bibnamefont{Dahlen}},
  \bibinfo{journal}{Phys. Rev. B} \textbf{\bibinfo{volume}{79}},
  \bibinfo{pages}{245306} (\bibinfo{year}{2009}).

\bibitem[{\citenamefont{von Friesen et~al.}(2009)\citenamefont{von Friesen,
  Verdozzi, and Almbladh}}]{fva.2009}
\bibinfo{author}{\bibfnamefont{M.~P.} \bibnamefont{von Friesen}},
  \bibinfo{author}{\bibfnamefont{C.}~\bibnamefont{Verdozzi}}, \bibnamefont{and}
  \bibinfo{author}{\bibfnamefont{C.-O.} \bibnamefont{Almbladh}},
  \bibinfo{journal}{Phys. Rev. Lett.} \textbf{\bibinfo{volume}{103}},
  \bibinfo{pages}{176404} (\bibinfo{year}{2009}).

\bibitem[{\citenamefont{Balzer et~al.}(2010{\natexlab{a}})\citenamefont{Balzer,
  Bauch, , and Bonitz}}]{bbb.2010}
\bibinfo{author}{\bibfnamefont{K.}~\bibnamefont{Balzer}},
  \bibinfo{author}{\bibfnamefont{S.}~\bibnamefont{Bauch}}, , \bibnamefont{and}
  \bibinfo{author}{\bibfnamefont{M.}~\bibnamefont{Bonitz}},
  \bibinfo{journal}{Phys. Rev. A} \textbf{\bibinfo{volume}{81}},
  \bibinfo{pages}{022510} (\bibinfo{year}{2010}{\natexlab{a}}).

\bibitem[{\citenamefont{Balzer et~al.}(2010{\natexlab{b}})\citenamefont{Balzer,
  Bauch, , and Bonitz}}]{bbb2.2010}
\bibinfo{author}{\bibfnamefont{K.}~\bibnamefont{Balzer}},
  \bibinfo{author}{\bibfnamefont{S.}~\bibnamefont{Bauch}}, , \bibnamefont{and}
  \bibinfo{author}{\bibfnamefont{M.}~\bibnamefont{Bonitz}},
  \bibinfo{journal}{Phys. Rev. A} \textbf{\bibinfo{volume}{82}},
  \bibinfo{pages}{033427} (\bibinfo{year}{2010}{\natexlab{b}}).

\bibitem{shbv.2015}  
N. Schl\"unzen, S. Hermanns, M. Bonitz, C. Verdozzi, cond-mat/arXiv:1508.02947.    
  
\bibitem{lr.2015}  
Y. Bar Lev and D. R. Reichman, cond-mat/arXiv:1508.05391.  
  
\bibitem[{\citenamefont{Ranitovic et~al.}(2011)\citenamefont{Ranitovic, Tong,
  Hogle, Zhou, Liu, Toshima, Murnane, and Kapteyn}}]{rthzetal.2011}
\bibinfo{author}{\bibfnamefont{P.}~\bibnamefont{Ranitovic}},
  \bibinfo{author}{\bibfnamefont{X.~M.} \bibnamefont{Tong}},
  \bibinfo{author}{\bibfnamefont{C.~W.} \bibnamefont{Hogle}},
  \bibinfo{author}{\bibfnamefont{X.}~\bibnamefont{Zhou}},
  \bibinfo{author}{\bibfnamefont{Y.}~\bibnamefont{Liu}},
  \bibinfo{author}{\bibfnamefont{N.}~\bibnamefont{Toshima}},
  \bibinfo{author}{\bibfnamefont{M.~M.} \bibnamefont{Murnane}},
  \bibnamefont{and} \bibinfo{author}{\bibfnamefont{H.~C.}
  \bibnamefont{Kapteyn}}, \bibinfo{journal}{Phys. Rev. Lett.}
  \textbf{\bibinfo{volume}{106}}, \bibinfo{pages}{193008}
  (\bibinfo{year}{2011}).

\bibitem[{\citenamefont{Pfeiffer and Leone}(2012)}]{pl.2012}
\bibinfo{author}{\bibfnamefont{A.~N.} \bibnamefont{Pfeiffer}} \bibnamefont{and}
  \bibinfo{author}{\bibfnamefont{S.~R.} \bibnamefont{Leone}},
  \bibinfo{journal}{Phys. Rev. A} \textbf{\bibinfo{volume}{85}},
  \bibinfo{pages}{053422} (\bibinfo{year}{2012}).

\bibitem[{\citenamefont{Chen et~al.}(2012)\citenamefont{Chen, Bell, Beck,
  Mashiko, Wu, Pfeiffer, Gaarde, Neumark, Leone, and Schafer}}]{cbbmetal.2012}
\bibinfo{author}{\bibfnamefont{S.}~\bibnamefont{Chen}},
  \bibinfo{author}{\bibfnamefont{M.~J.} \bibnamefont{Bell}},
  \bibinfo{author}{\bibfnamefont{A.~R.} \bibnamefont{Beck}},
  \bibinfo{author}{\bibfnamefont{H.}~\bibnamefont{Mashiko}},
  \bibinfo{author}{\bibfnamefont{M.}~\bibnamefont{Wu}},
  \bibinfo{author}{\bibfnamefont{A.~N.} \bibnamefont{Pfeiffer}},
  \bibinfo{author}{\bibfnamefont{M.~B.} \bibnamefont{Gaarde}},
  \bibinfo{author}{\bibfnamefont{D.~M.} \bibnamefont{Neumark}},
  \bibinfo{author}{\bibfnamefont{S.~R.} \bibnamefont{Leone}}, \bibnamefont{and}
  \bibinfo{author}{\bibfnamefont{K.~J.} \bibnamefont{Schafer}},
  \bibinfo{journal}{Phys. Rev. A} \textbf{\bibinfo{volume}{86}},
  \bibinfo{pages}{063408} (\bibinfo{year}{2012}).

\bibitem[{\citenamefont{Chen et~al.}(2013)\citenamefont{Chen, Wu, Gaarde, and
  Schafer}}]{cwgs.2013}
\bibinfo{author}{\bibfnamefont{S.}~\bibnamefont{Chen}},
  \bibinfo{author}{\bibfnamefont{M.}~\bibnamefont{Wu}},
  \bibinfo{author}{\bibfnamefont{M.~B.} \bibnamefont{Gaarde}},
  \bibnamefont{and} \bibinfo{author}{\bibfnamefont{K.~J.}
  \bibnamefont{Schafer}}, \bibinfo{journal}{Phys. Rev. A}
  \textbf{\bibinfo{volume}{87}}, \bibinfo{pages}{033408}
  (\bibinfo{year}{2013}).

\bibitem[{\citenamefont{Wu et~al.}(2013)\citenamefont{Wu, Chen, Gaarde, and
  Schafer}}]{wcgs.2013}
\bibinfo{author}{\bibfnamefont{M.}~\bibnamefont{Wu}},
  \bibinfo{author}{\bibfnamefont{S.}~\bibnamefont{Chen}},
  \bibinfo{author}{\bibfnamefont{M.~B.} \bibnamefont{Gaarde}},
  \bibnamefont{and} \bibinfo{author}{\bibfnamefont{K.~J.}
  \bibnamefont{Schafer}}, \bibinfo{journal}{Phys. Rev. A}
  \textbf{\bibinfo{volume}{88}}, \bibinfo{pages}{043416}
  (\bibinfo{year}{2013}).

\bibitem[{\citenamefont{Chini et~al.}(2014)\citenamefont{Chini, Wang, Cheng,
  and Chang}}]{cwcc.2014}
\bibinfo{author}{\bibfnamefont{M.}~\bibnamefont{Chini}},
  \bibinfo{author}{\bibfnamefont{X.}~\bibnamefont{Wang}},
  \bibinfo{author}{\bibfnamefont{Y.}~\bibnamefont{Cheng}}, \bibnamefont{and}
  \bibinfo{author}{\bibfnamefont{Z.}~\bibnamefont{Chang}}, \bibinfo{journal}{J.
  Phys. B: At. Mol. Opt. Phys.} \textbf{\bibinfo{volume}{47}},
  \bibinfo{pages}{124009} (\bibinfo{year}{2014}).

\bibitem[{smi()}]{smiles1}
\bibinfo{note}{J. Fern\'andez Rico, I. Ema, R. L\'opez, G. Ram\'irez and K.
  Ishida, in {\em Recent Advances in Computational Chemistry: Molecular
  Integrals over Slater Orbitals}, eds. T. Ozdogan and M. B. Ruiz (Transworld
  Research Network, 2008), pp. 145.}

\bibitem[{\citenamefont{Rico et~al.}(2004)\citenamefont{Rico, Lopez, Ramirez,
  and Ema}}]{smiles2}
\bibinfo{author}{\bibfnamefont{J.~F.} \bibnamefont{Rico}},
  \bibinfo{author}{\bibfnamefont{R.}~\bibnamefont{Lopez}},
  \bibinfo{author}{\bibfnamefont{G.}~\bibnamefont{Ramirez}}, \bibnamefont{and}
  \bibinfo{author}{\bibfnamefont{I.}~\bibnamefont{Ema}}, \bibinfo{journal}{J.
  Comput. Chem.} \textbf{\bibinfo{volume}{25}}, \bibinfo{pages}{1987}
  (\bibinfo{year}{2004}).

\bibitem[{\citenamefont{Liao et~al.}(2015)\citenamefont{Liao, Sandhu, Camp,
  Schafer, and Gaarde}}]{lscsg.2015}
\bibinfo{author}{\bibfnamefont{C.-T.} \bibnamefont{Liao}},
  \bibinfo{author}{\bibfnamefont{A.}~\bibnamefont{Sandhu}},
  \bibinfo{author}{\bibfnamefont{S.}~\bibnamefont{Camp}},
  \bibinfo{author}{\bibfnamefont{K.~J.} \bibnamefont{Schafer}},
  \bibnamefont{and} \bibinfo{author}{\bibfnamefont{M.~B.}
  \bibnamefont{Gaarde}}, \bibinfo{journal}{Phys. Rev. Lett.}
  \textbf{\bibinfo{volume}{114}}, \bibinfo{pages}{143002}
  (\bibinfo{year}{2015}).

\bibitem[{\citenamefont{Bunge et~al.}(1993)\citenamefont{Bunge, Barrientos, and
  Bunge}}]{bbb.1993}
\bibinfo{author}{\bibfnamefont{C.~F.} \bibnamefont{Bunge}},
  \bibinfo{author}{\bibfnamefont{J.~A.} \bibnamefont{Barrientos}},
  \bibnamefont{and} \bibinfo{author}{\bibfnamefont{A.~V.} \bibnamefont{Bunge}},
  \bibinfo{journal}{Atomic Data and Nuclear Data Tables}
  \textbf{\bibinfo{volume}{53}}, \bibinfo{pages}{113} (\bibinfo{year}{1993}).

\bibitem[{not()}]{notepeakpos}
\bibinfo{note}{Similarly to He case the absolute position of the peaks is
  shifted with respect to experiment}.

\bibitem[{\citenamefont{S\"akkinen et~al.}(2012)\citenamefont{S\"akkinen,
  Manninen, and van Leeuwen}}]{smvl.2012}
\bibinfo{author}{\bibfnamefont{N.}~\bibnamefont{S\"akkinen}},
  \bibinfo{author}{\bibfnamefont{M.}~\bibnamefont{Manninen}}, \bibnamefont{and}
  \bibinfo{author}{\bibfnamefont{R.}~\bibnamefont{van Leeuwen}},
  \bibinfo{journal}{New J. Phys.} \textbf{\bibinfo{volume}{14}},
  \bibinfo{pages}{013032} (\bibinfo{year}{2012}).

\end{thebibliography}
\end{document}